\documentclass[twocolumn,pre,aps,showpacs,amsmath,amssymb]{revtex4-1}
\usepackage{graphicx}
\usepackage{bm}
\usepackage{soul}
\usepackage{color}

\begin{document}

\title{Controlling fingering instabilities in Hele-Shaw flows in the presence of wetting film effects}
\author{Pedro H. A. Anjos$^{1}$}
\email[]{pamorimanjos@iit.edu}
\author{M. Zhao$^{2}$}
\email[]{mzhao9@uci.edu}
\author{J. Lowengrub$^{2}$}
\email[]{lowengrb@math.uci.edu}
\author{Weizhu Bao$^{3}$}
\email[]{matbaowz@nus.edu.sg}
\author{Shuwang Li$^{1}$}
\email[]{sli@math.iit.edu}
\affiliation{$^{1}$ Department of Applied Mathematics, Illinois Institute of Technology,
	Chicago, Illinois  60616, USA\\ 
	$^2$ Department of Mathematics, University of California Irvine, Irvine, California 92697, USA\\
$^{3}$ Department of Mathematics, National University of Singapore, Singapore, Singapore 119067, Singapore}


\begin{abstract}

In this paper, the interfacial motion between two immiscible viscous fluids in the confined geometry of a Hele-Shaw cell is studied. We consider the influence of a thin wetting film trailing behind the displaced fluid, which dynamically affects the pressure drop at the fluid-fluid interface by introducing a nonlinear dependence on the interfacial velocity. In this framework, two cases of interest are analyzed: The injection-driven flow (expanding evolution), and the lifting plate flow (shrinking evolution). In particular, we investigate the possibility of controlling the development of fingering instabilities in these two different Hele-Shaw setups when wetting effects are taken into account. By employing linear stability theory, we find the proper time-dependent injection rate $Q(t)$ and the time-dependent lifting speed ${\dot b}(t)$ required to control the number of emerging fingers during the expanding and shrinking evolution, respectively. Our results indicate that the consideration of wetting leads to an increase in the magnitude of $Q(t)$ [and ${\dot b}(t)$] in comparison to the non-wetting strategy. Moreover, a spectrally accurate boundary integral approach is utilized to examine the validity and effectiveness of the controlling protocols at the fully nonlinear regime of the dynamics and confirms that the proposed injection and lifting schemes are feasible strategies to prescribe the morphologies of the resulting patterns in the presence of the wetting film.

\end{abstract}
\maketitle

\section{Introduction}
\label{intro}

The Saffman-Taylor instability~\cite{PG} arises at the interface separating two viscous fluids constrained to flow in the narrow gap between closely spaced parallel plates of an effectively two-dimensional (2D) Hele-Shaw cell. In its radial geometry setup~\cite{Lp}, the Saffman-Taylor instability occurs when a fluid is injected through a central inlet, displacing a higher viscosity fluid radially outwards. As the initially almost circular fluid-fluid interface expands, it deforms, and fingerlike protuberances form. The perturbed interface evolves, and the produced fingers split at their tips, ultimately leading to the formation of highly branched interfacial patterns presenting fingers of different lengths~\cite{homsy1987viscous,casademunt2004viscous,mccloud1995experimental}. In this way, one can say that in the injection-driven, constant-gap radial fingering instability the most emblematic pattern formation processes associated with it are finger ramification and finger competition.

A different type of fingering instability arises if the top plate of the Hele-Shaw cell is lifted
in the direction perpendicular to the cell plates~\cite{Ben2,Roche,Shelley1,Roy,Anke,Tarafdar,Ben3,Anke2,Tarafdar2,Nase,Diasmaster,Stone}. This lifting Hele-Shaw cell problem is a variant of the classical constant-gap Hele-Shaw situation in which the cell gap width is time-dependent. In the lifting case, initially one has an almost circular interface separating a more viscous fluid surrounded by a less viscous one. Then, while the lower cell plate is held fixed, the upper plate is lifted parallel to it, establishing a time-dependent gap flow. During the lifting process, the outer fluid enters the gap through the sides and displaces the inner fluid, making the interface unstable due to the Saffman-Taylor instability. As a consequence, the shrinking interface deforms as the fingers of the outer fluid invade the inner fluid. Eventually, the amplitude of the perturbations increases, and longer, smooth, competing fingers of the outer fluid move towards the center of the cell. Note that in this situation, the instabilities are driven by the variable gap width, and not by injection of fluid.   

In many industrial applications, the development of fingering instabilities may be undesirable. One emblematic example in which the emergence of interfacial instabilities is very unwelcome is during oil recovery~\cite{Gorell,Stokes}, where petroleum is displaced by injection of water into the oil field in an attempt to extract more oil from the well. Depending on how this process is conducted, rapidly evolving ramified fingers may bypass the oil in the reservoir and reach the point of extraction, and thus leading to poor oil recovery. It is also known that viscous fingering
has a potentially harmful character in applications involving adhesives~\cite{Anke,Nase,pedroadh}. So, processes aimed towards minimizing the fingering instabilities, or controlling the growth of viscous fingers are of technological and scientific importance.

One of the methods to control the development of interfacial instabilities consists of using specific time-dependent injection fluxes $Q(t)$ (in the case of expanding flow) and gap widths $b(t)$ (for shrinking flow). For example, in the framework of an injection-driven radial Hele-Shaw flow performed under $Q(t)\sim t^{-1/3}$, it has been demonstrated by fully nonlinear simulations and experiments that the system evolves into well behaved $n$-fold symmetric, self-similar structures~\cite{ShuwangPRL}. Even though this particular process was not able to eliminate the interfacial disturbances, it does offer a valid way to prescribe and control the morphology of the resulting patterns, avoiding
the appearance of inconvenient branched morphologies (formed by finger ramification) that arise when using the usual constant injection rate $Q$~\cite{homsy1987viscous,casademunt2004viscous,mccloud1995experimental}. As for lifting Hele-Shaw flows, the simulations performed under a variable gap $b(t)\sim t^{-2/7}$ in Ref.~\cite{Zhao18} revealed the emergence of $n$-fold morphologies that do not vanish as the interface shrinks though the evolution is not self-similar. 

Despite the relatively large number of investigations of minimization and controlling schemes for Hele-Shaw problems~\cite{mama1,mama2,mama3,mama4,ShuwangPRL,Zhao18,Stone,Liam,Juel,Stone1,Bon}, none of these studies consider the influence of wetting films on the dynamics of the interfacial evolution. However, depending on the nature of the fluids involved they can wet the walls of the Hele-Shaw cell plates, leaving behind a film of finite thickness.
In fact, wetting is the ability of a liquid to maintain contact with a 
solid surface, resulting from intermolecular interactions when the 
two are brought together. It deals with three phases of 
matter: gas, liquid, and solid. Wetting is an ubiquitous phenomena along with
the interface dynamics in fluid mechanics and materials science \cite{Gennes1,Gennes2,Starov,Thompson}. 
Thus it is very important to include wetting effects in modeling and simulation of
interface dynamics in fluid mechanics and materials science \cite{QianSheng,Ren2010,Bao2012,Bao2017}. 

In a seminal paper~\cite{Park} Park and Homsy have shown that the consideration of such wetting effects leads to nonnegligible corrections in the pressure difference at the fluid-fluid interface, introducing a nonlinear dependence on the interfacial velocity. A number of subsequent theoretical and experimental investigations in rectangular Hele-Shaw cells~\cite{Tab,Schwartz,Saf2,Rei1,Rei2} have indicated that the inclusion of wetting effects helps to provide a better match between theory and experiments. This has also been the case for injection-driven flows in the radial Hele-Shaw cell setup~\cite{Max,Russo,Max_amp,pedrowet,jackson} and for time-dependent gap flows in the lifting Hele-Shaw arrangement~\cite{Pedro,pedroadh}. In particular, by employing a weakly nonlinear analysis, the authors of Ref.~\cite{pedrowet} have demonstrated that the inclusion of wetting effects can significantly impact fingering formation at the onset of nonlinearities, providing overall stabilization of the fingers, and restraining the development of both finger bifurcation and finger length variability. Later on, these weakly nonlinear findings were confirmed by fully nonlinear simulations in Ref.~\cite{jackson}. In addition, in Ref.~\cite{jackson} the authors found that the number of fingers produced in the first ramification of the interface is generally different to that when no dynamic wetting is included.

From what we have discussed in the previous paragraphs, it is clear that the wetting film plays a major role in the interfacial development in Hele-Shaw flows, and its effect cannot be neglected. Even though Refs.~\cite{jackson,Pedro,pedrowet} have done a good job in exploring the impact of wetting in Hele-Shaw flows, these studies focused mainly on analyzing morphological aspects of the patterns that arise in these systems, rather than developing controlling schemes. Motivated by these points, our main purpose in this work is to perform a theoretical investigation of controlling protocols that properly takes into account the effects of this thin film for two different types of Hele-Shaw flows. In particular, by considering the presence of the unavoidable wetting thin film, we seek to design feasible and more accurate controlling strategies that could be utilized for technological and industrial purposes. 

Our study initiates by utilizing linear theory to determine the time-dependent injection strategy $Q(t)$ that keeps the mode of largest growth rate $n_{\rm max}$ unmodified as the interface evolves. In contrast to the studies that have utilized similar time-dependent approaches~\cite{Stone,Zhao18,ShuwangPRL,mama1,Juel}, our protocol is carefully designed to account for the thin wetting film left behind by the displaced fluid. Then, we use a boundary integral formalism to verify the efficiency of this linear-stability-based injection protocol in controlling the development of viscous fingering instabilities during fully nonlinear stages of the dynamics. We also apply similar maneuvering to design a time-dependent lifting speed ${\dot b}(t)$ capable of controlling fingering formation in the lifting Hele-Shaw cell setup.

The remainder of this paper is organized as follows. In Section~\ref{gov} we present the governing equations for the injection-driven Hele-Shaw flow together with a derivation of the linear growth rate for the system taking into account wetting effects. The boundary integral scheme utilized to gain access to the interfacial shapes is demonstrated in Sec~\ref{num}. In Sec.~\ref{self} we present our proposed controlling injection scheme and analyze its efficiency in the expanding evolution of the interface at fully nonlinear regime. Following a similar approach, in Section~\ref{derivation2} we discuss the shrinking evolution subjected to our controlling time-dependent lifting speed. Our final conclusions are compiled in Sec.~\ref{conclude}.

\begin{figure}[t]
	\includegraphics[width=2.6 in]{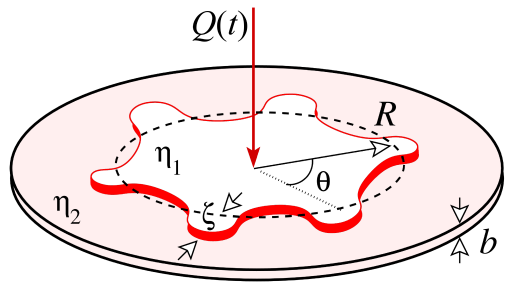}
	\caption{Representative sketch of the injection-driven flow in a radial Hele-Shaw cell.}
	\label{geom1}
\end{figure}

\section{INJECTION-DRIVEN HELE-SHAW FLOW}
\label{derivation}

\subsection{Governing equations and linear growth rate}
\label{gov}

Consider a Hele-Shaw cell of constant gap spacing $b$ containing two immiscible, incompressible, and viscous fluids (see Fig.~\ref{geom1}). Denote the viscosities of the inner and outer fluids, respectively, as $\eta_{1}$ and $\eta_{2}$. Between the two fluids there exists a surface tension $\sigma$. Fluid 1 is injected into fluid 2 at a given injection rate $Q=Q(t)$, which may depend on time. We describe the
perturbed fluid-fluid interface as ${\cal R}(\theta,t)= R(t) + \zeta(\theta,t)$, where
$\theta$ represents the azimuthal angle and $R(t)$ is the time-dependent unperturbed radius,
\begin{equation}
\label{radius}
R(t)=\sqrt{R_{0}^{2} + \frac{1}{\pi} \int_{0}^{t} Q(t')dt'},
\end{equation}
with $R_{0}$ being the unperturbed radius at $t=0$. The presence of the time integral in Eq.~(\ref{radius}) is required since the injection rate is not necessarily constant. In addition,  $\zeta(\theta,t)=\sum_{n=-\infty}^{+\infty} \zeta_{n}(t) \exp{(i n \theta)}$ denotes the net interface perturbation with Fourier amplitudes $\zeta_{n}(t)$, and discrete wave numbers $n$. Our main task in this section is to obtain the linear growth rate of interfacial
perturbations.

For the effectively two-dimensional geometry of the radial Hele-Shaw cell, the governing equation of the system is the gap-averaged Darcy's law~~\cite{PG,homsy1987viscous,mccloud1995experimental,casademunt2004viscous,Lp}
\begin{equation}
\label{Darcy1}
{\textbf v}_{j} =-\frac{b^{2}}{12 \eta_{j}}\bm{\nabla}p_{j},
\end{equation}
where ${\textbf v}_{j}$ and $p_{j}$ denote the velocity and pressure in fluids $j=1, 2$, respectively. From the irrotational nature of the flow 
(${\bm \nabla} \times {\bf v}_{j}=0$), and the incompressibility condition 
\begin{equation}
\label{incompress}
{\bm \nabla}\cdot {\bf v}_{j}=0, 
\end{equation}
it can be readily verified that the velocity potential $\phi_{j}$ obeys Laplace equation ${\bm{\nabla}}^2 \phi_{j}=0$. It this context, to 
get the equation of motion for the interface, we rewrite~(\ref{Darcy1}) for each of the fluids in terms of the velocity potential. Integrate 
and then subtract the resulting equations from each other to obtain
\begin{equation}
\label{difference}
\left(\frac{\beta-1}{\beta+1}\right) \left ( \frac{\phi_1 + \phi_2}{2}\right ) - \left ( \frac{\phi_1 - 
	\phi_2}{2} \right ) = - \frac{b^{2}(p_{1} - p_{2})}{12(\eta_{1} + \eta_{2})},
\end{equation}
where the dimensionless parameter $\beta=\eta_{2}/\eta_{1}$ is the viscosity ratio.

To include the contributions coming from surface tension and wetting effects we consider a generalized Young-Laplace pressure 
boundary condition, which expresses the pressure jump across the fluid-fluid interface~\cite{Bret,Park,Tab,Schwartz,Saf2,Rei1,Rei2,Russo,Max_amp,enric_wet}
\begin{equation}
\label{pressure}
p_{1}-p_{2}= \frac{\pi}{4} \sigma \kappa + \frac{2 \sigma}{b} \cos{\alpha_c} \left [1  + J_{0}  {{\rm Ca}_{l}}^{2/3} \right ].
\end{equation}
The first term on the right-hand side (RHS) of Eq.~(\ref{pressure}) represents the contribution related to surface tension and the interfacial curvature $\kappa$ in the plane of the Hele-Shaw cell. The factor $\pi /4$ is purely a capillary static effect, coming from the $z$-average of the mean interfacial curvature. The second term on the RHS of Eq.~(\ref{pressure}) accounts for the 
contribution of the constant curvature associated with the interface profile in the direction perpendicular to the Hele-Shaw cell plates, set by the static contact angle $\alpha_{c}$ measured between the plates and the curved meniscus. As in most experiments and wetting models, we consider a nonwetting fluid (fluid 1) displacing a wetting one (fluid 2), so that $\alpha_{c}=0$. The second term also considers the effect of a thin wetting film trailing behind the displaced fluid, where ${\rm Ca}_{l}=\eta_{2} V/ \sigma$ is the local capillary number, $V$ the normal component of the interface velocity, and $J_{0}=3.8$. Note that this term is crucial to this paper, and it has been originally proposed by a prior theoretical work by Park and Homsy~\cite{Park}. They were the first to conduct such a theoretical analysis by combining Bretherton's lubrication approximation~\cite{Bret} with the Saffman-Taylor equations~\cite{homsy1987viscous}, via double asymptotic expansion of the ratio of film thickness to transverse characteristic length and capillary number raised to 1/3. Following Park and Homsy's analysis existing experimental and theoretical results were reconciled, thereby elucidating the important role of wetting film in the nonlinear finger formation process. Equation (\ref{pressure}) imposes a dynamic boundary condition via the local capillary number ${\rm Ca}_{l}$, which is natural in interface dynamics in fluid mechanics and materials science \cite{Barrat99,Qian08,Liu19}.

The problem is then specified by the generalized pressure jump boundary condition~(\ref{pressure}), plus the kinematic boundary condition which states 
that the normal components of each fluid's velocity are continuous at the interface
\begin{equation}
\label{kine}
{\bf n} \cdot \bm{\nabla} \phi_{1} = {\bf n} \cdot\bm{\nabla} \phi_{2},
\end{equation}
with ${\bf n}$ representing the unit normal vector at the interface.

Following standard steps performed in previous perturbative studies for Hele-Shaw flows~\cite{pedrowet}, first, we define Fourier expansions 
for the velocity potentials. Then, we express $\phi_{j}$ in terms of the perturbation amplitudes $\zeta_n$ by considering condition~(\ref{kine}). 
Substituting these relations, and the pressure jump condition Eq.~(\ref{pressure}) into Eq.~(\ref{difference}), always keeping terms up to 
first-order in $\zeta$, and Fourier transforming, we find the dimensionless equation of motion for the perturbation amplitudes 
\begin{eqnarray}
\label{result}
\dot{\zeta}_{n}=\lambda(n) \zeta_{n}, 
\end{eqnarray}
where the overdot denotes total time derivative,
\begin{eqnarray}
\label{growth}
\lambda(n)&=&\frac{1}{1 + w(n)} \Bigg [ \frac{Q}{2 \pi R^{2}} \left(\frac{\beta-1}{\beta+1}  |n| - 1 \right ) \nonumber\\
&& -\frac{\pi}{4 {\rm Ca}_{g} R^{3}} \frac{\beta}{(\beta+1)} |n| (n^{2} - 1) \Bigg ],
\end{eqnarray}
is the linear growth rate, and
\begin{equation}
\label{w}
w(n)= |n| J_{0} \frac{\beta}{(\beta+1)} \frac{1}{9qR} \left ( \frac{24 \pi R q^2}{Q {\rm Ca}_{g}} \right )^{1/3}
\end{equation}
is related to the wetting film contribution. Here, lengths and time are rescaled by $R_{0}$, and ${R_0}^2/Q_{0}$, respectively, and $Q_{0}$ is the injection rate at $t=0$. The global capillary number 
\begin{equation}
\label{cap}
{\rm Ca}_{g} = \frac{12 \eta_{2} R_{0} Q_{0} }{\sigma b^2}
\end{equation}
expresses a relative measure of viscous to surface tension forces, while $q = R_{0}/b$ is the initial aspect ratio. 

The expressions~(\ref{result})-(\ref{w}) represent the linear equations of the viscous fingering problem in a radial 
Hele-Shaw cell, taking into consideration the contributions from wetting film effects. We have verified that by setting $J_{0}=0$, Eqs.~(\ref{result})-(\ref{w}) reproduce the equivalent expressions originally derived in Ref.~\cite{Mir4}, where the 
effects of the wetting film are not taken into account. Note that, in this limit, a proper match with their results is obtained if the extra $\pi /4$ multiplicative 
factor appearing in the surface tension term in Eq.~(\ref{pressure}) is replaced by one.

\subsection{Numerical scheme}
\label{num}

Using potential theory~\cite{book}, the solution to the Laplace equation can be written in terms of boundary integrals. Considering the Darcy's law written in terms of $\phi_{j}$, we take $\displaystyle \phi_1=-\beta{p_1}$ and $\displaystyle \phi_2=-p_2$ to be dimensionless potential functions of fluid 1 and fluid 2, respectively. These potential functions are harmonic and have continuous normal derivatives across the interface. Thus, the velocity potential $\phi$ satisfies a double layer potential
\begin{equation}
\phi(\mathbf{x})=\frac{1}{2\pi}\int_{\Gamma}\gamma(\mathbf{x}') \left(\frac{\partial \ln |\mathbf{x}-\mathbf{x}'|}{\partial \mathbf{n(x')}}+1 \right)d s(\mathbf{x'})+\frac{Q}{2 \pi}\ln |\mathbf{x}|,\label{exphi}
\end{equation}
where $\gamma(\mathbf{x})$ is the dipole density on the interface $\Gamma$ and $\mathbf{x}$ denotes
the position vector with the origin located at the center of the cell. 

Using the dimensionless kinematic boundary condition and the pressure jump across the interface, we obtain
\begin{eqnarray}
&~&~\frac{1}{2}\left(\frac 1 \beta+1\right)\gamma(\mathbf{x}) \nonumber\\
&+&\frac{1}{2\pi}\left(\frac 1\beta-1\right) \bigg [ \bigg. \int_{\Gamma}\gamma(\mathbf{x}')\left(\frac{\partial \ln |\mathbf{x}-\mathbf{x}'|}{\partial \mathbf{n(x')}}+1\right)d s(\mathbf{x'})\nonumber\\
&+& Q\ln |\mathbf{x}|\bigg. \bigg ] =\frac{1}{{\rm Ca}_{g}}\left[2q(1+J_0|{\rm Ca}_l|^{2/3})+\frac{\pi}{4}\kappa\right],\label{eqmu}
\end{eqnarray}
and once $\gamma(\mathbf{x})$ is solved, we are able to compute the normal velocity of the interface as
\begin{equation}
V(\mathbf{x})=\frac{1}{2\pi}\int_{\Gamma}\gamma_{s}(\mathbf{x}')\frac{(\mathbf{x}-\mathbf{x}')^\perp\cdot\mathbf{n(x)}}{|\mathbf{x}-\mathbf{x}'|^2}ds(\mathbf{x'})+ \frac{Q}{2 \pi}\frac{\mathbf{x}\cdot\mathbf{n}}{|\mathbf{x}|^2},\label{V1}
\end{equation}
where the subscript $s$ denotes the partial derivative with respect to arclength $s$ and ${\textbf{x}}^{\perp}=(x_2,-x_1)$. 

Equation~(\ref{eqmu}) is a Fredholm integral equation of second-kind, and it is coupled with Eq.~(\ref{V1}) via   $\displaystyle {\rm Ca}_l=\frac{\eta_2Q_0}{\sigma R_0}V$. We apply an iterative method to solve it, and $V_n$, the normal velocity of the interface at time $t_n$, is needed to compute the local capillary number ${\rm Ca}_l$. While it is not known a prior, we use an initial guess from the previous time step, i.e., $V_n^0=V_{n-1}$ and $V_1^0=0$ for the first step. Thus, we are able to compute ${\rm Ca}_l$ and solve Eq. (\ref{eqmu}) via GMRES~\cite{GMRES}. Then we compute the normal velocity utilizing Eq.~(\ref{V1}) and use a Picard iteration~\cite{jackson} to generate the normal velocity for the next Picard step $k$. The normal velocity is updated as 
\begin{equation}
{V}_n^k=V_n^{k-1}+\psi(\tilde{V}_n^k-V_n^{k-1}),
\end{equation}
where $\tilde{V}_n^k$ represents the value of Eq.(\ref{V1}) at time $t_n$, Picard step $k$, and $\psi$ is a relaxation coefficient. Once the new velocity is obtained, ${\rm Ca}_l$ can be computed and the process is repeated until the updated normal velocity does not change from the previous Picard step. That is, the error between updated velocities is less than a tolerance. Here we set the tolerance to be $10^{-9}$ which ensures that the normal velocity is accurate enough for Eq.(\ref{eqmu}), especially for a long time simulation such as the ones performed in this work. The relaxation coefficient $\psi$ highly depends on ${\rm Ca}_l$ and the shape of the interface. We take $\psi$ to be $0.01\sim 0.1$ for large ${\rm Ca}_l$ and complicated initial shapes (i.e., more modes and larger perturbation). It takes about $10\sim 100$ steps for the Picard iteration to converge, and therefore long-time simulations are very expensive. To overcome this issue, in Appendix A we introduce a rescaling idea~\cite{ShuwangJCP,Zhao17} that improves the efficiency of our numerical method. 

Once the normal velocity $V(\mathbf{x})$ is determined, the interface is evolved through
\begin{equation}
\label{nonexp}
\frac{d {\mathbf{x}}}{d {t}}\cdot \mathbf{n}=V(\mathbf{x}).
\end{equation}
Note that this system is very stiff due to the higher-order terms introduced by the curvature and requires a severe third-order time-step constraint $\Delta t\sim h^3$, where $\Delta t$ is the time step, and $h$ is the spatial grid size. Following the small scale-decomposition~\cite{HLS}, we remove the stiffness and obtain a second-order accurate updating scheme in time.

\subsection{Expanding evolution with fixed number of fingers}
\label{self}

In this section, our task is to determine what is the functional form of a time-dependent injection rate $Q(t)$ for which the number of fingers remains unchanged as time progresses. Recall that, at the linear level, an estimate for the number of fingers formed during the injection process is given by the closest integer to the mode of largest growth rate $n_{\rm max}$~\cite{Lp,homsy1987viscous,mccloud1995experimental,casademunt2004viscous}. Therefore, the desired $Q(t)$ will be the one that keeps $n_{\rm max}$ unmodified as the interface evolves.

It is clear from Eq.~(\ref{radius}) that the unperturbed radius $R(t)$ satisfies 
\begin{equation}
\label{Revol}
\dot R = \frac{Q(t)}{2 \pi R(t)},
\end{equation}
while the Fourier amplitudes evolve as predicted by the linear Eq.~(\ref{result}). To characterize the interface morphology and quantitatively measure how much the fluid-fluid interface deviates from the reference circle, we introduce the rescaled perturbation mode amplitudes $\zeta_{n}/R = \zeta_{n}(t)/R(t)$, whose evolution is given by
\begin{eqnarray}
\label{SF}
\frac{d}{dt} \left(\frac{\zeta_{n}}{R}\right) = \Lambda(n)  \left(\frac{\zeta_{n}}{R}\right),
\end{eqnarray}
where
\begin{eqnarray}
\label{modlam}
\Lambda(n)=\frac{\dot{\zeta}_{n}}{\zeta_{n}} - \frac{\dot R}{R}=\lambda(n) - \frac{Q}{2 \pi R^2}
\end{eqnarray}
is the modified linear growth rate.

\begin{figure}[t]

	\includegraphics[scale=0.55]{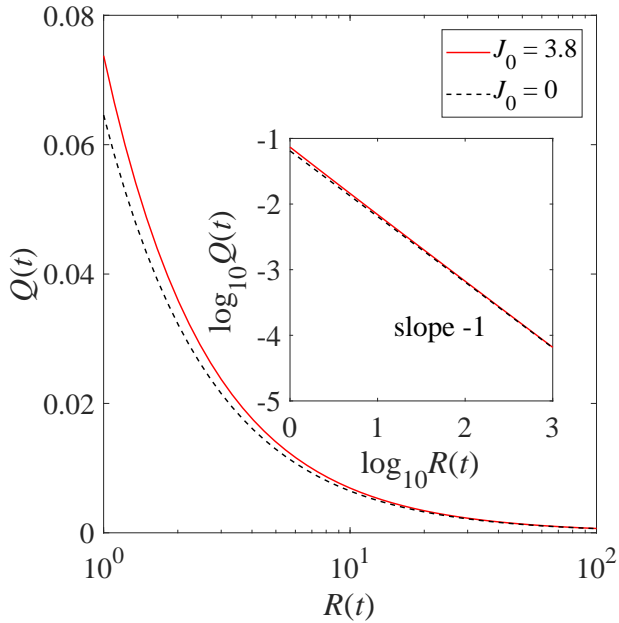}
	
	\caption{Time-dependent injection rate $Q(t)$ as a function of $R(t)$, for the cases with ($J_{0}=3.8$), and without wetting ($J_{0}=0$). The solid curve corresponds to the flux $Q_{n_{\rm max}}(t)$, while the dashed one is obtained by considering Eq.~(\ref{Qnowet}). Here $n_{\rm max} = 5$, $\beta=10$, $q=25$, and ${\rm Ca}_{g} = 1000$.}
	\label{Qcomp}
\end{figure}

\begin{figure*}[t]

	\includegraphics[scale=0.51]{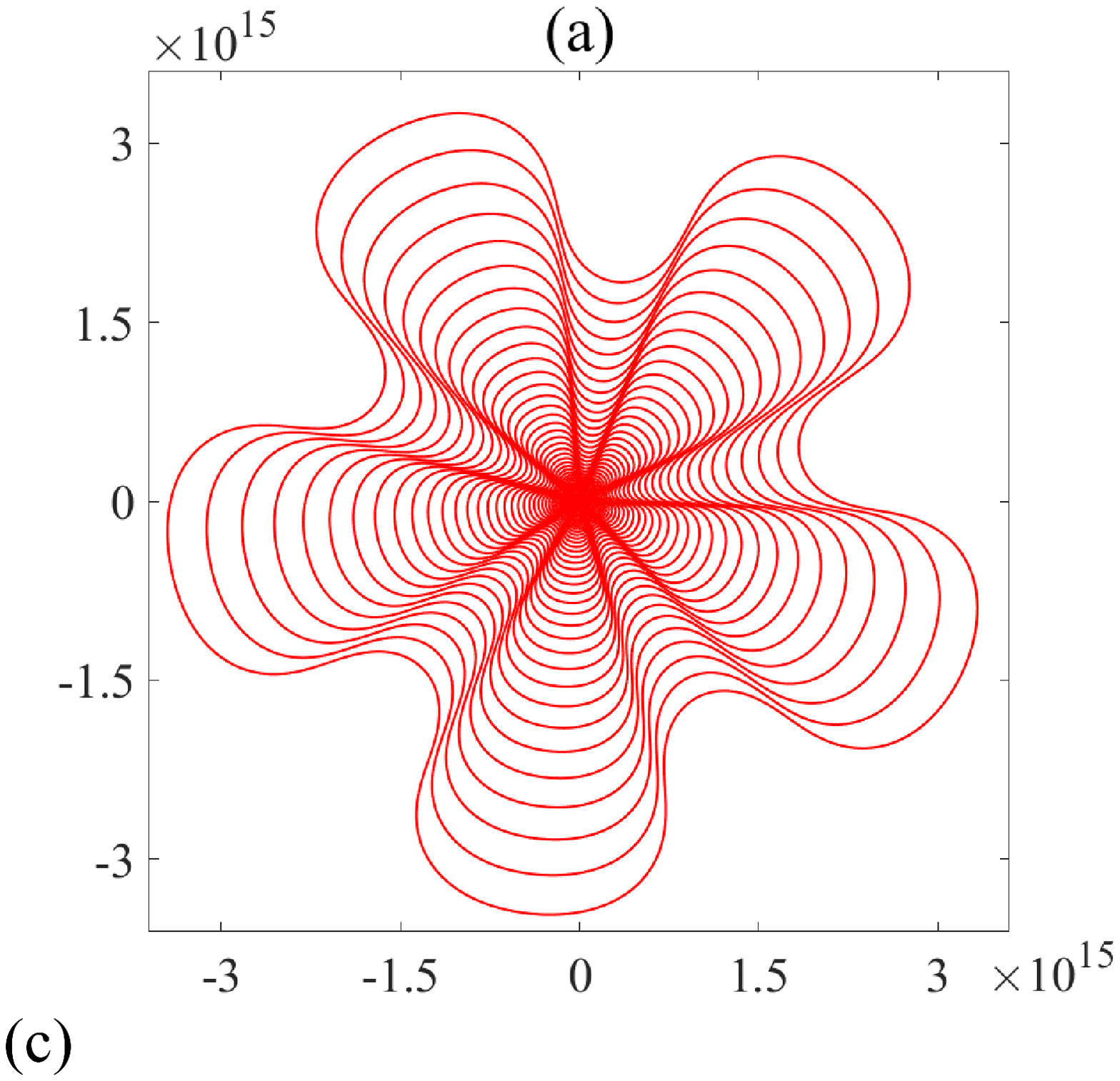}
	\includegraphics[scale=0.5]{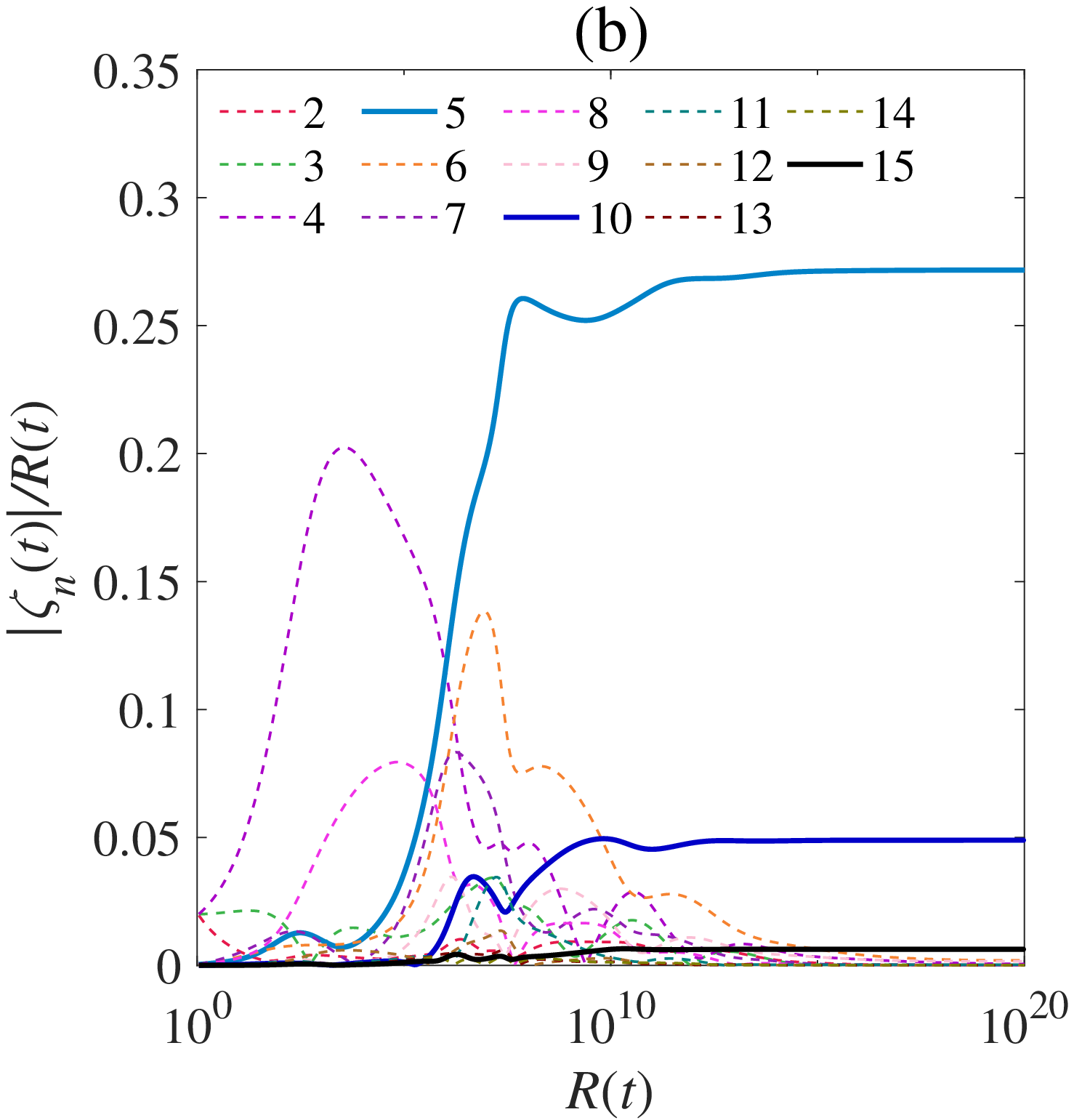}\\
	\includegraphics[scale=0.38]{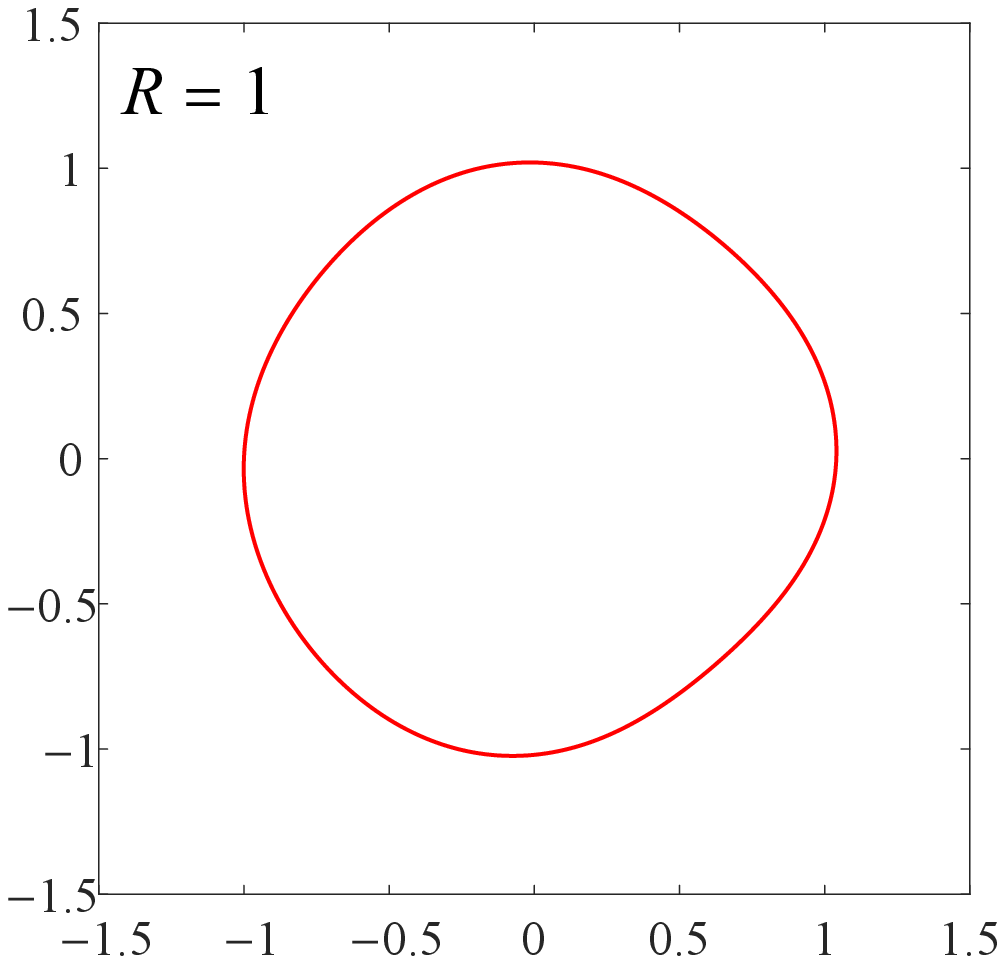}
	\includegraphics[scale=0.38]{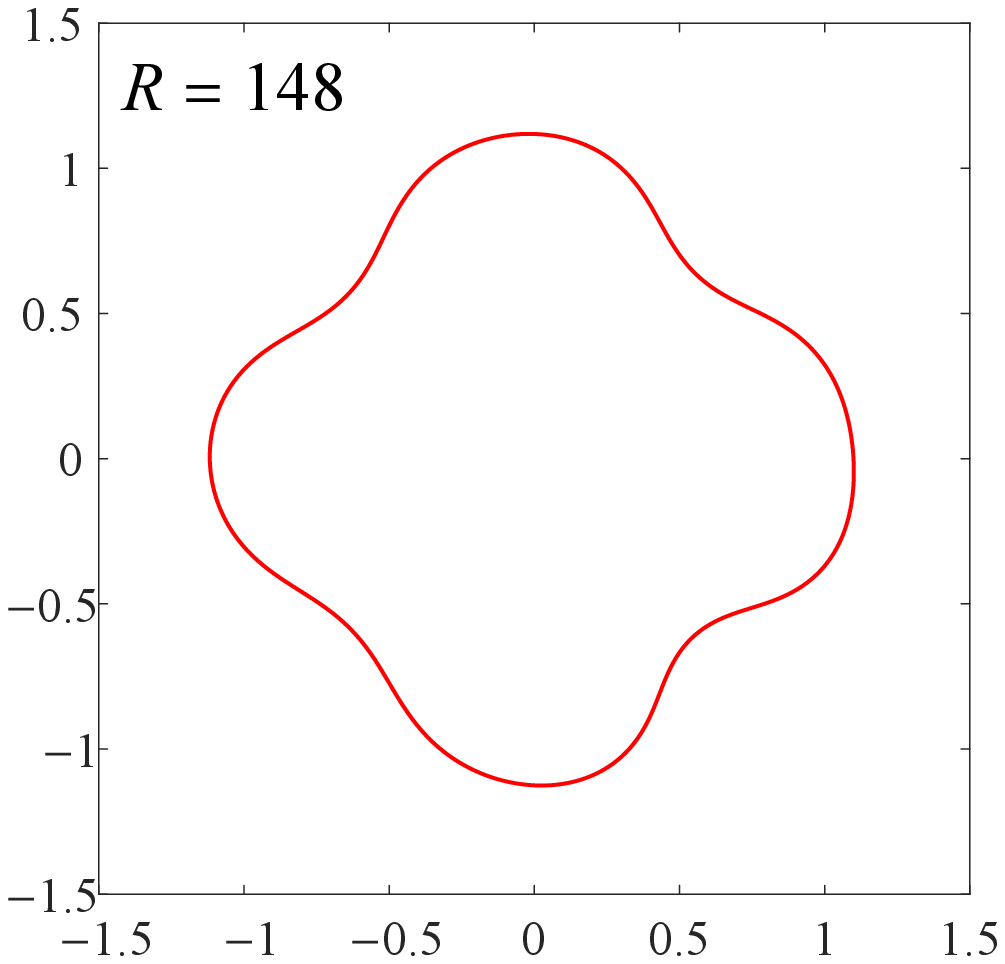}
	\includegraphics[scale=0.38]{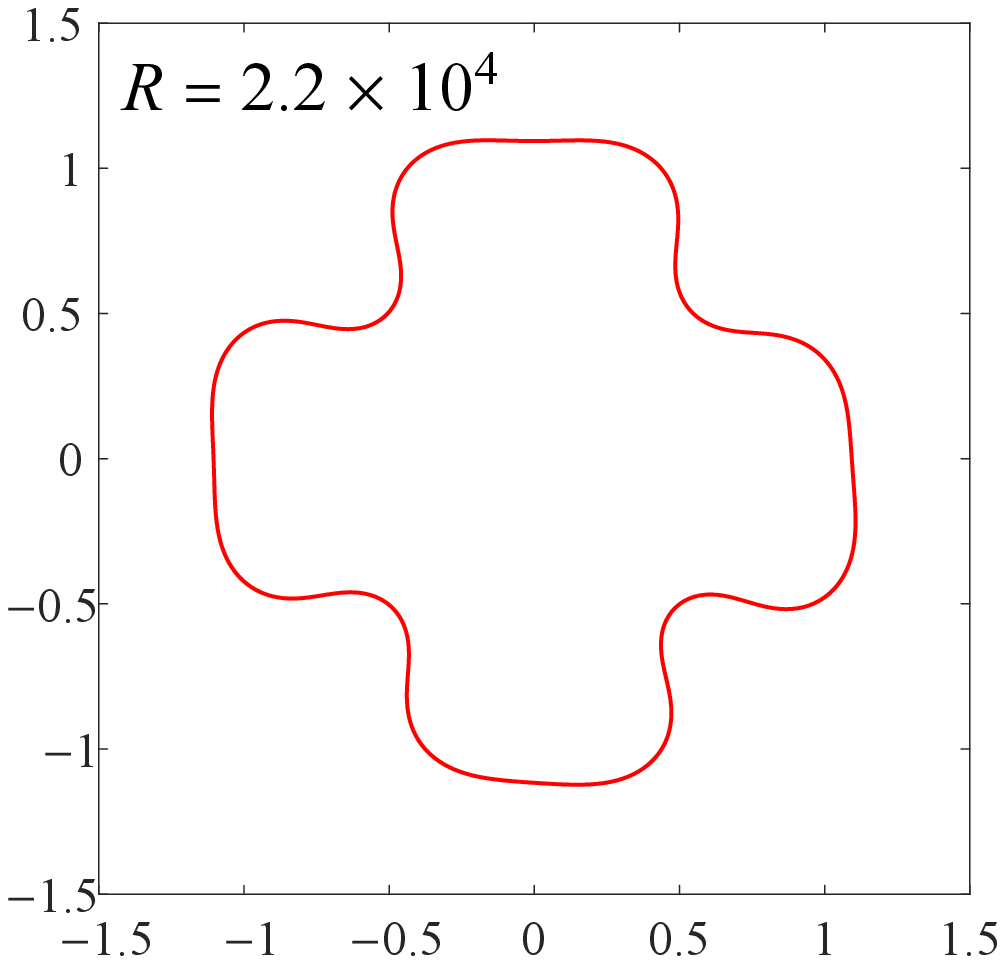}
	\includegraphics[scale=0.38]{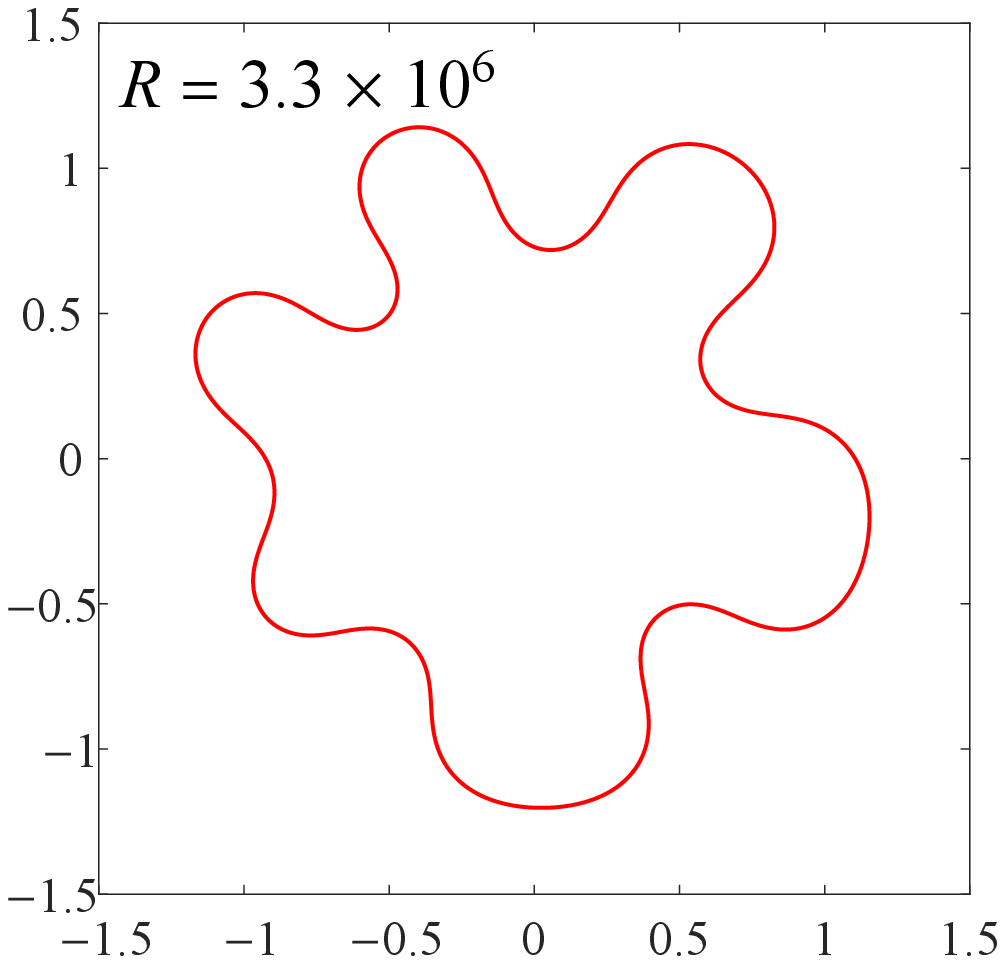}\\
	\includegraphics[scale=0.38]{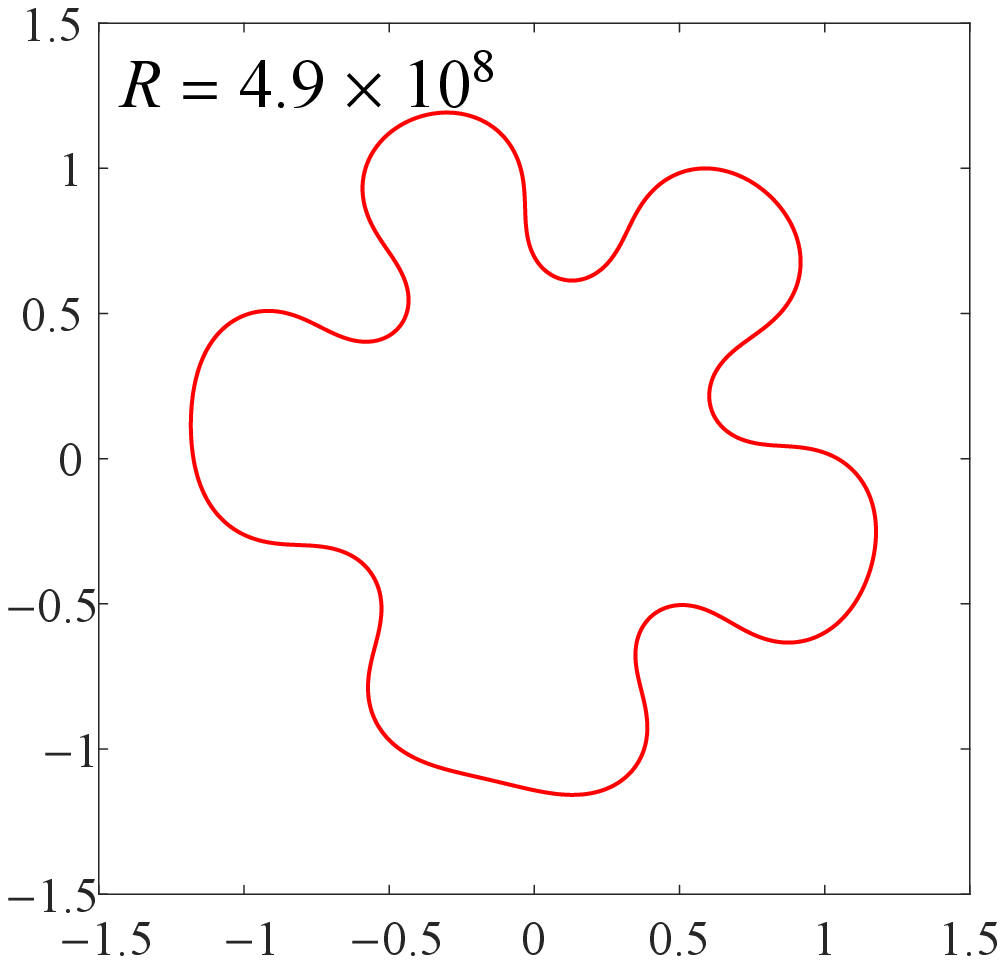}
	\includegraphics[scale=0.38]{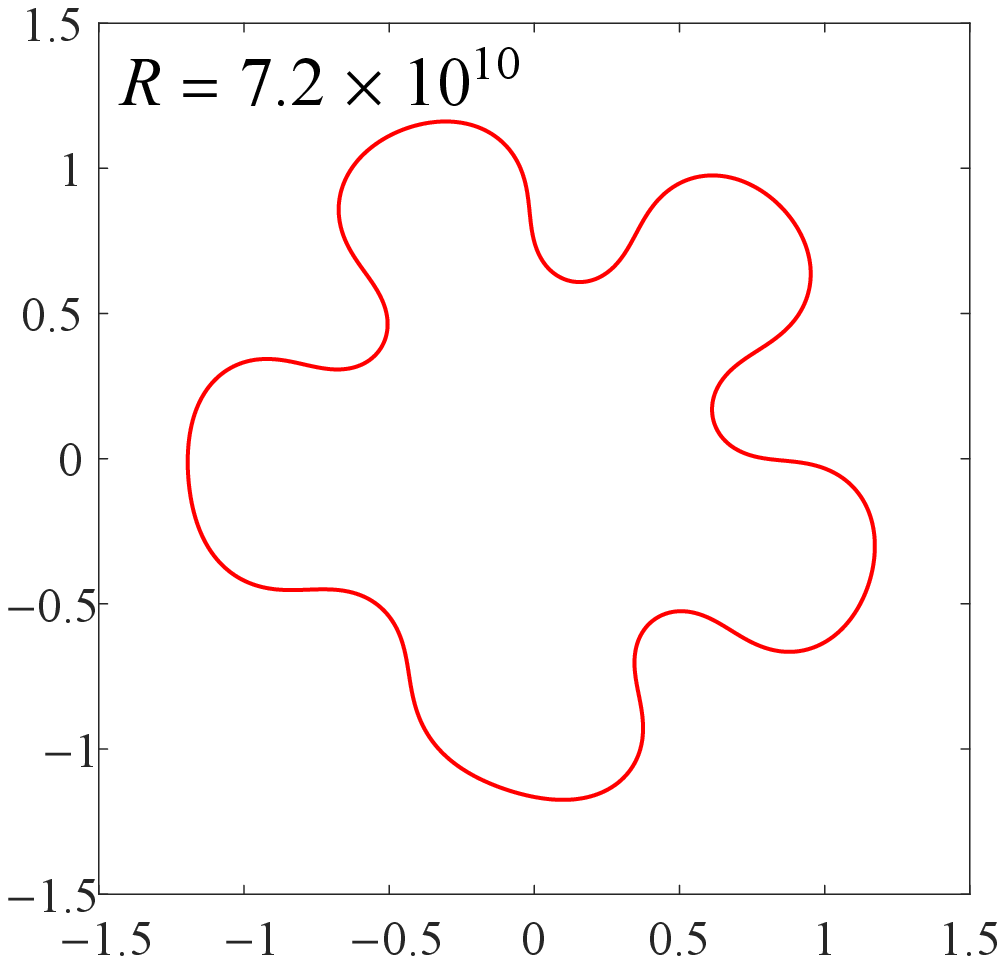}
	\includegraphics[scale=0.38]{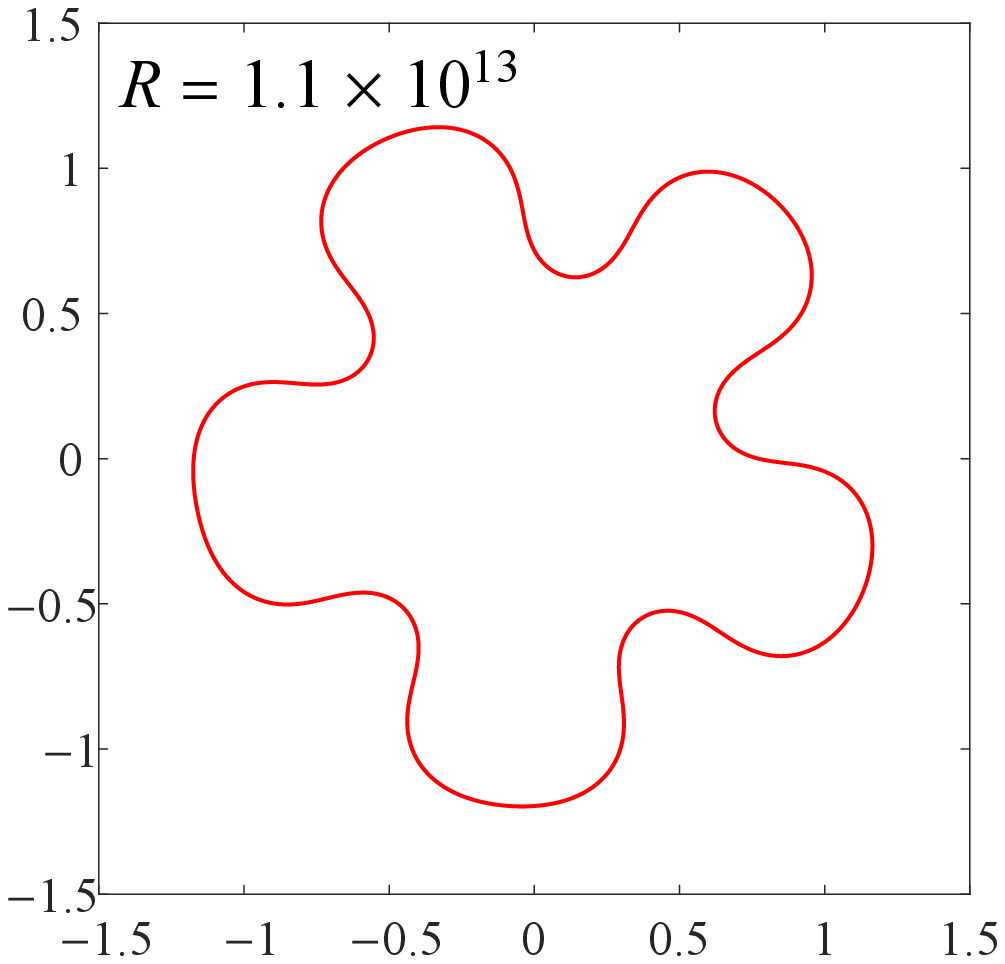}
	\includegraphics[scale=0.38]{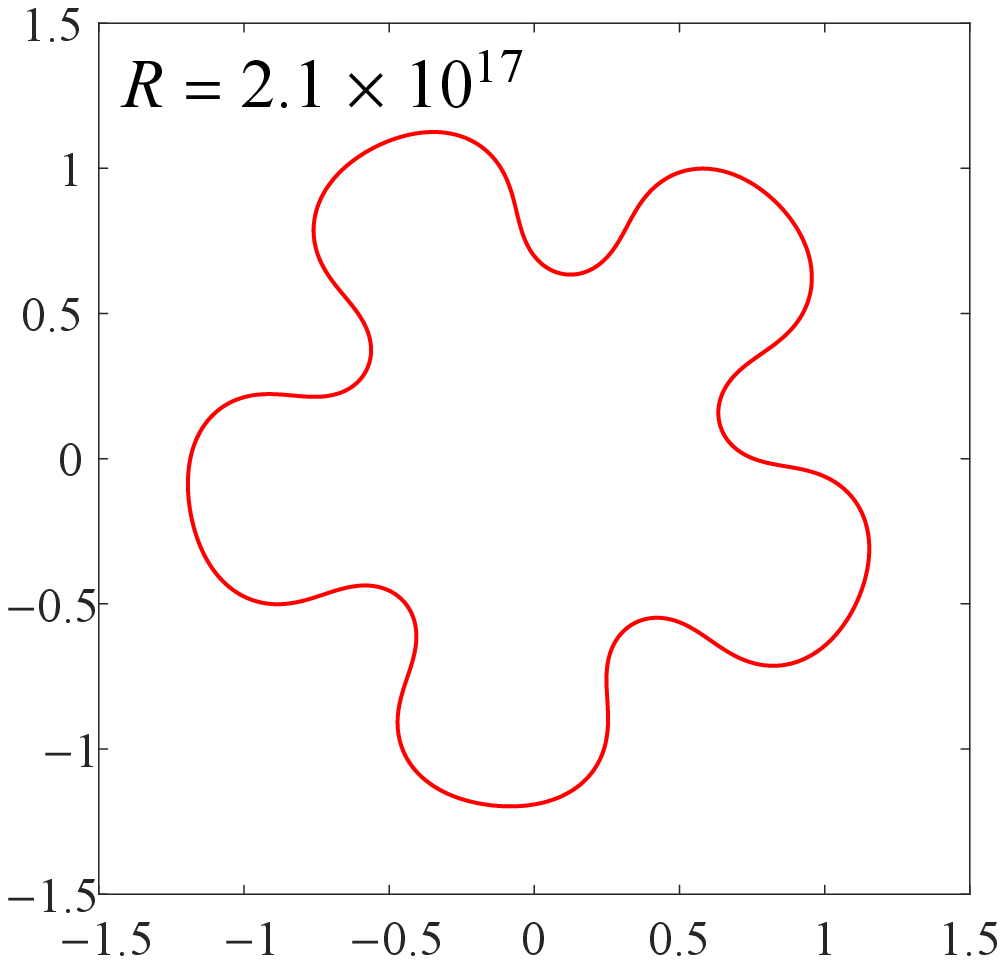}
	
	\caption{Interface dynamics under the flux $Q_{n_{\rm max}}(t)$ with $n_{\rm max} = 5$. In (a) we depict a time overlaid plot of the evolving interface shown at equal time intervals, and (b) illustrates the corresponding fully nonlinear variation of the rescaled perturbation amplitudes $|\zeta_{n} (t)|/R(t)$ with $R(t)$ for all the modes in the interval $2 \leq n \leq 15$. In (c) we break the evolution depicted in (a) into snapshots of the interface at different values of unperturbed radius $R(t)$. Moreover, each plot frame in (c) is rescaled by its corresponding value of $R(t)$. Here $\beta=10$, $q=25$, ${\rm Ca}_{g} = 1000$, and the initial condition is ${\cal R}(\theta,0)=1+0.02(\sin2\theta+\cos3\theta+\cos4\theta)$.}
	\label{exp2}
\end{figure*}

Our control strategy is based on keeping the mode of largest growth rate $n_{\rm max}$ of the system, that is, the mode for which ${d\Lambda}/{dn}|_{n=n_{\rm max}}=0$, unchanged. Therefore, the desired $Q(t)$ to accomplish this task is found by setting the derivative of Eq.~(\ref{modlam}) equal to zero, and utilizing Eqs.~(\ref{growth}) and~(\ref{w}), we obtain after some algebra
\begin{eqnarray}
\label{SF3}
&~&\frac{1}{2\pi R^2}\left(\frac{\beta -1}{\beta+1} \right) Q \nonumber\\
&+& \frac{J_{0}\beta}{18\pi (\beta+1) q R^3} \left(\frac{24\pi R q^2}{{\rm Ca}_{g}} \right)^{1/3}Q^{2/3} -\frac{\pi \beta (3n_{\rm max}^2-1)}{4{\rm Ca}_{g} (\beta+1) R^3} \nonumber\\
&-& \frac{\pi J_{0} \beta^2 n_{\rm max}^3}{18 (\beta+1)^2 q R^4 {\rm Ca}_{g}} \left(\frac{24\pi R q^2}{{\rm Ca}_{g}} \right)^{1/3} Q^{-1/3} = 0.
\end{eqnarray}
Note that by defining $x \equiv Q^{1/3}$, we identify Eq.~(\ref{SF3}) as a quartic equation. It is possible to find all the solutions of Eq.~(\ref{SF3}) and write closed-form analytical expressions for $Q(t)$. However, in spite of the somewhat cumbersome nature of these closed-form solutions, we prefer to keep all the necessary information to solve Eq.~(\ref{SF3}) in Appendix B. In addition, the specific format of these solutions does not offer many physical insights.

\begin{figure*}
	
	\includegraphics[scale=0.45]{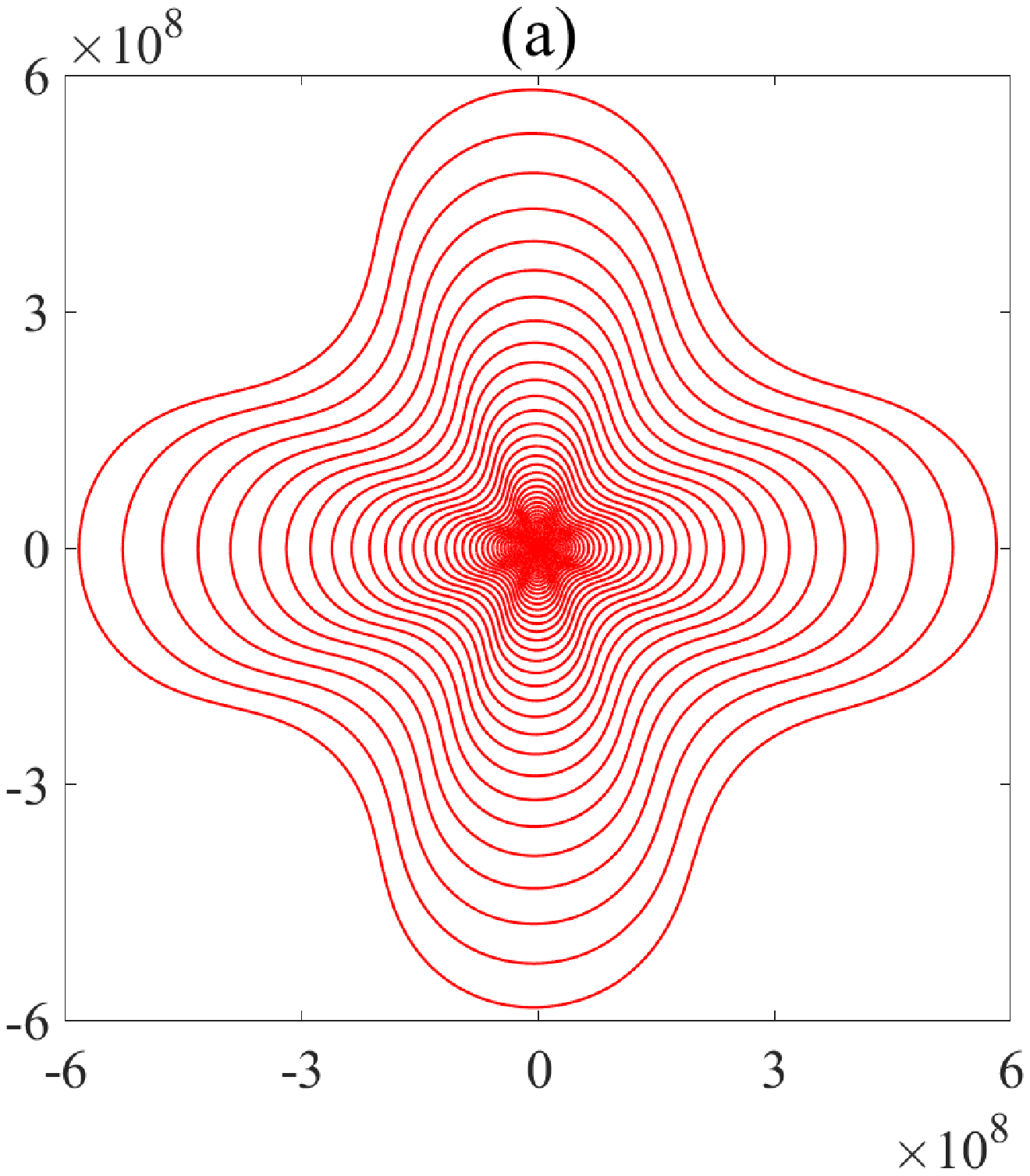}
	\includegraphics[scale=0.45]{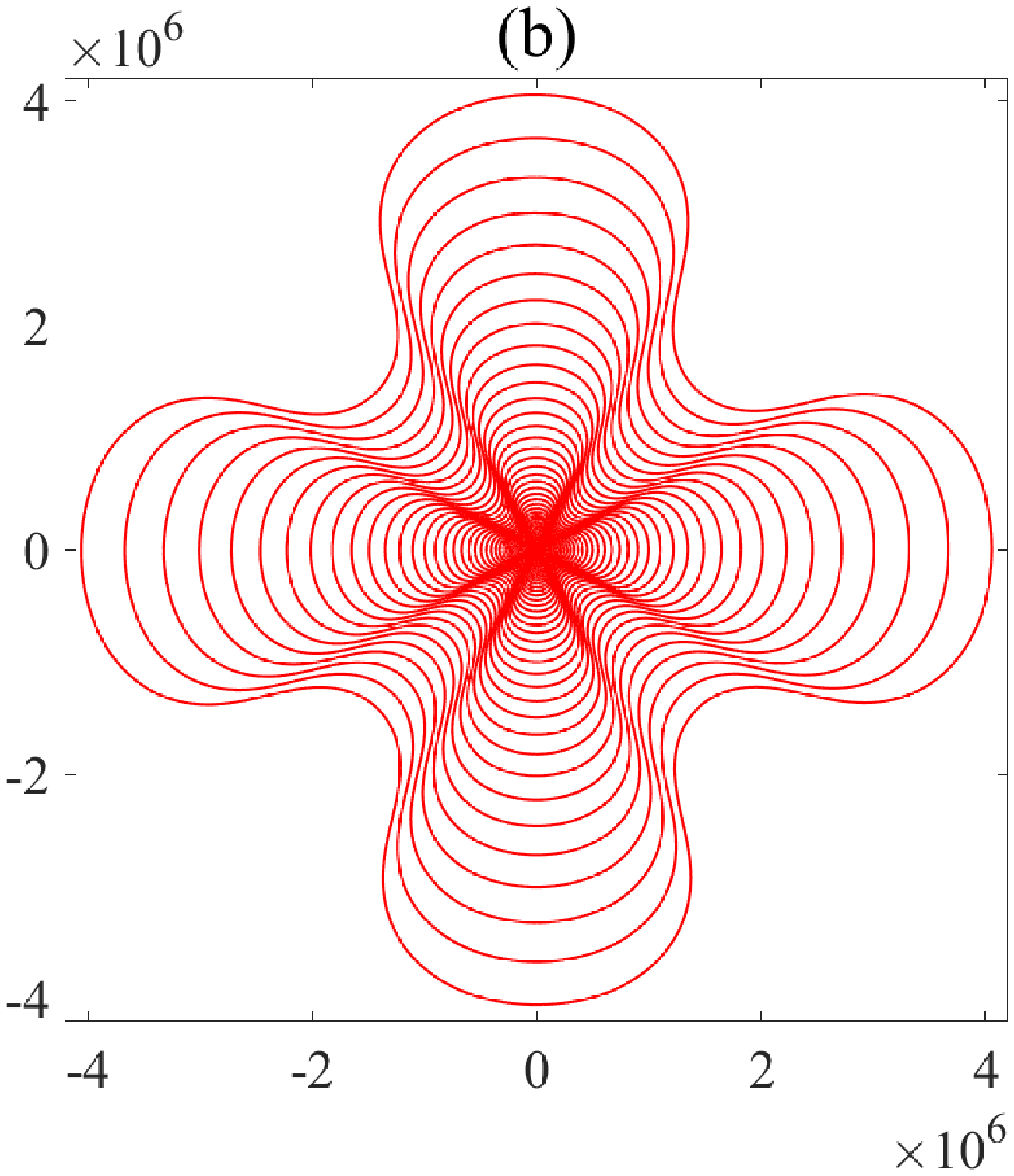}\\
	\includegraphics[scale=0.45]{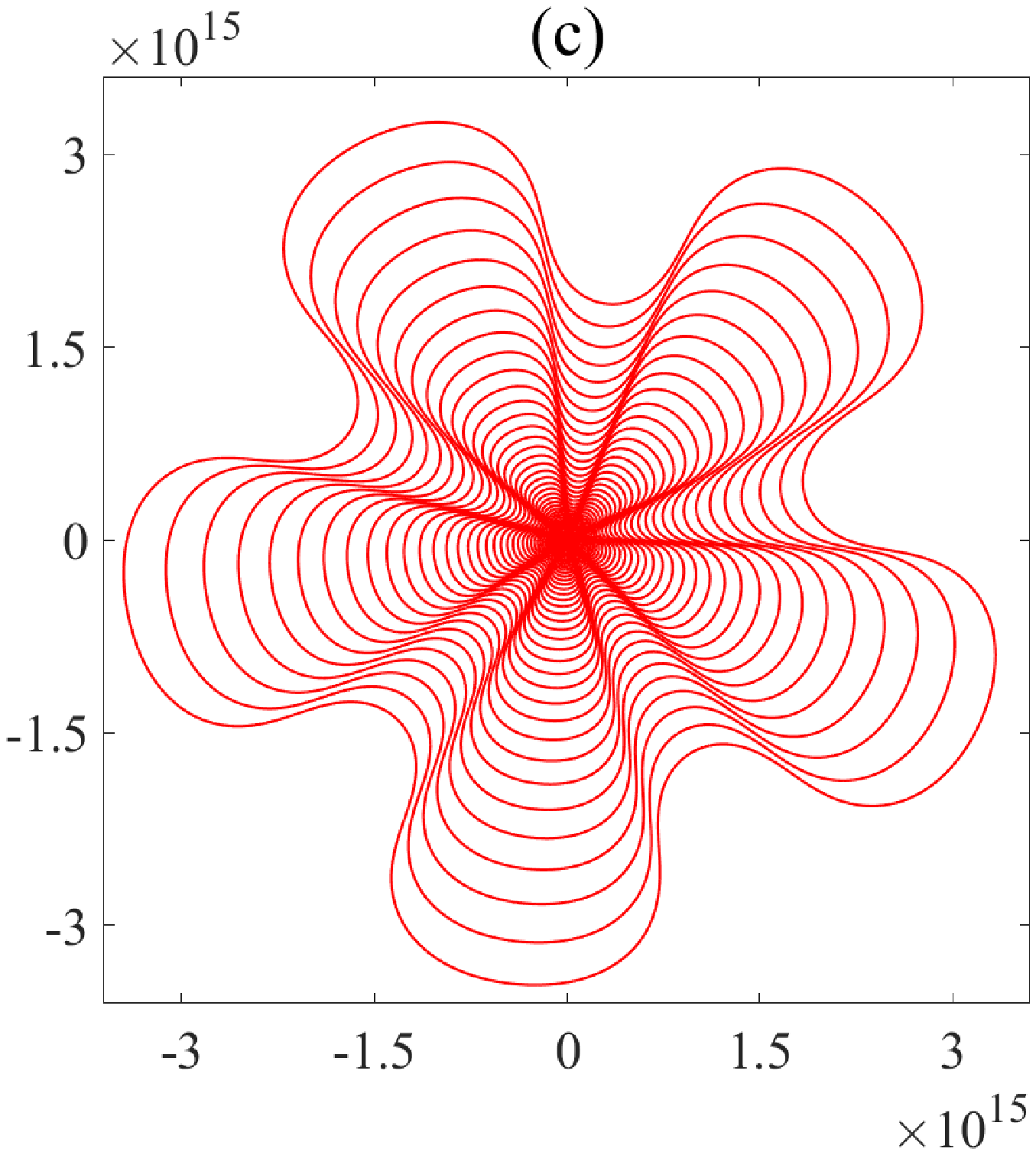}
	\includegraphics[scale=0.45]{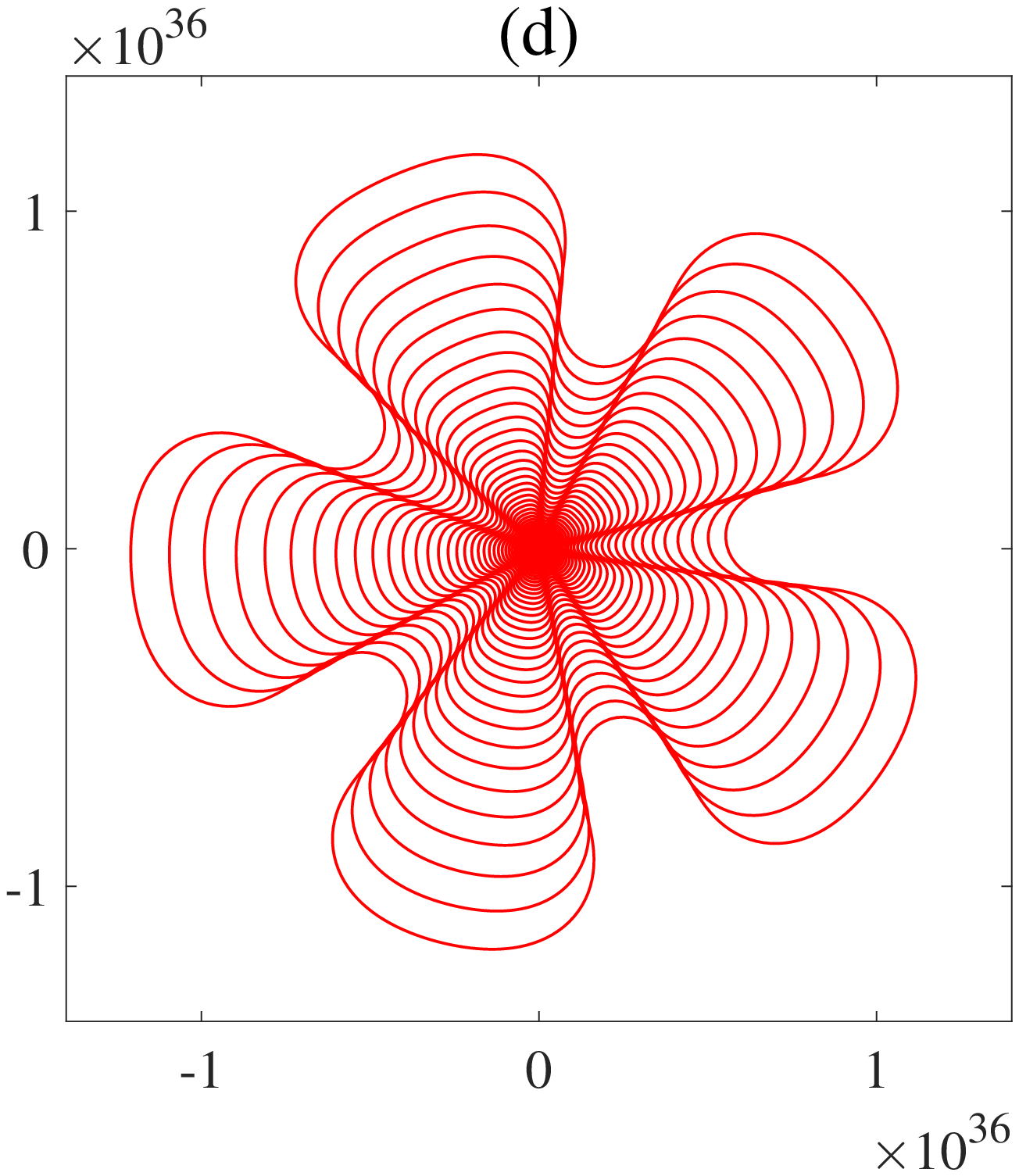}\\
	\includegraphics[scale=0.45]{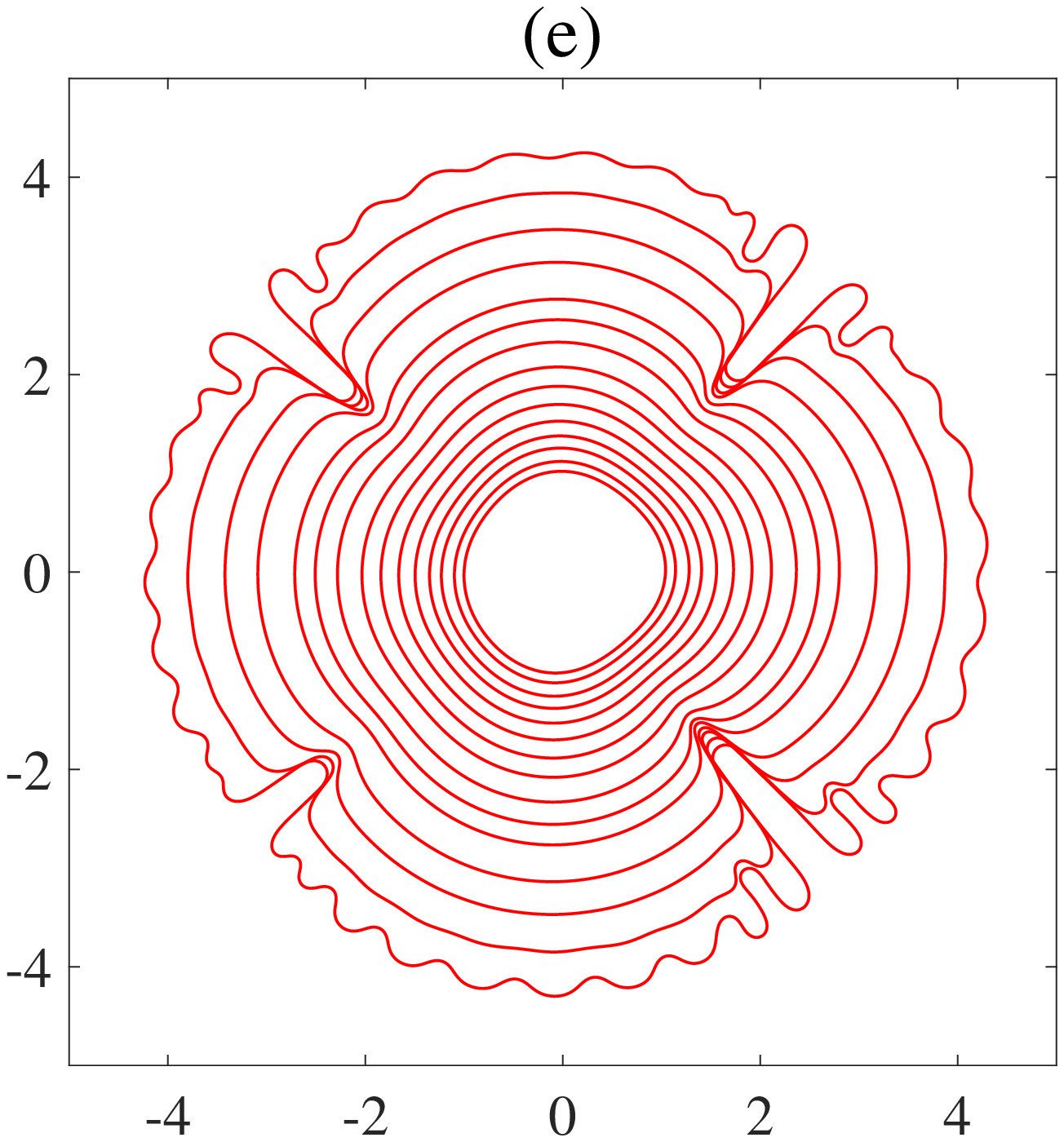}
	\includegraphics[scale=0.45]{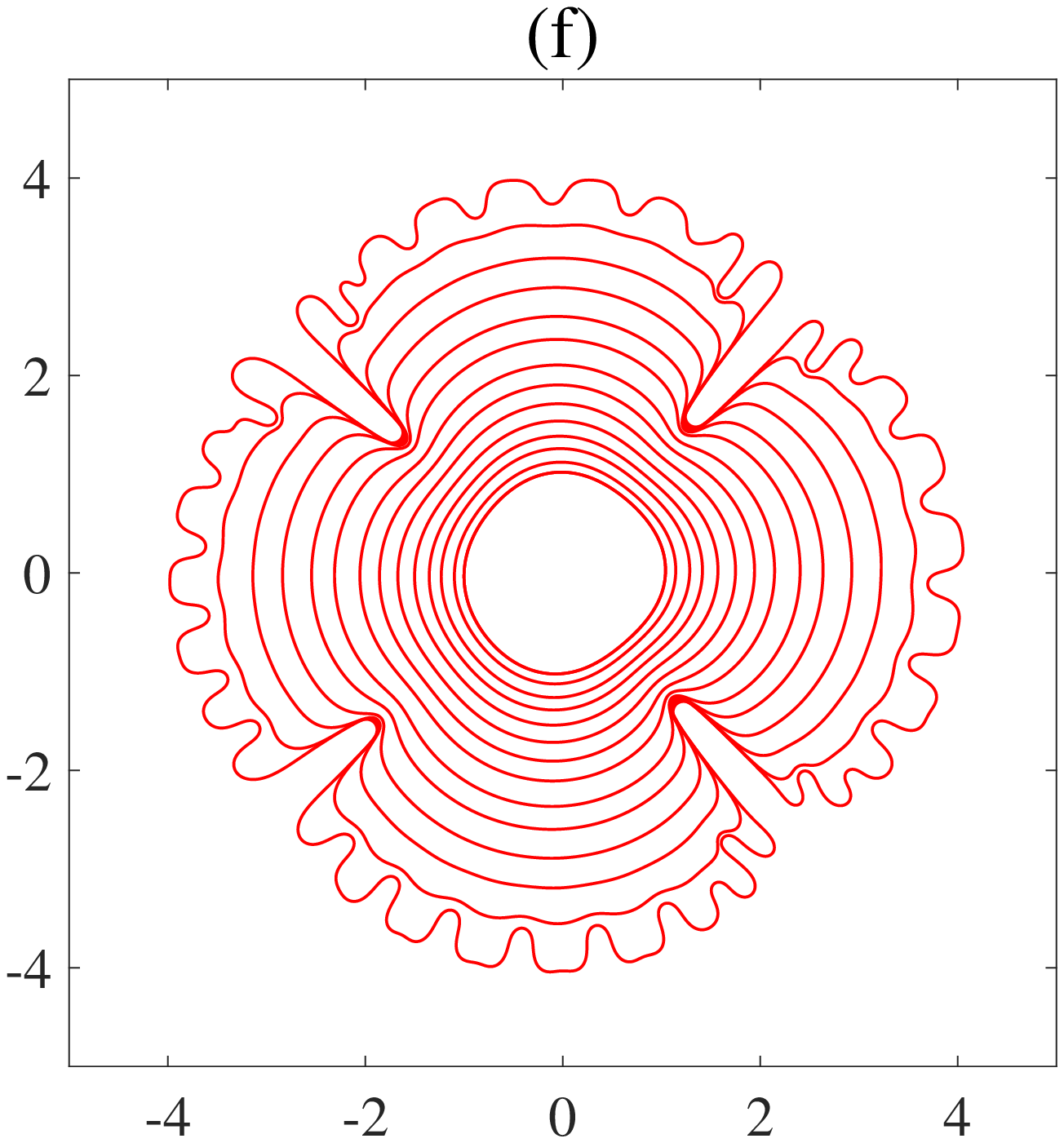}
	
	\caption{Time evolution of the fully nonlinear interfacial patterns produced by the time-dependent controlling injection rate $Q_{n_{\rm max}}(t)$ [(a)-(d)] obtained by solving Eq.~(\ref{SF3}), and by the usual constant injection [(e) and (f)] rate. In (a) and (b) [(c) and (d)] we impose that mode $n_{\rm max} = 4$ [$n_{\rm max} = 5$] is the fastest growing. The values for the viscosity ratio are: $\beta = 10$ (for all the patterns in the left column) and $\beta = 100$ (for all the patterns in the right column). All the other physical parameters and initial conditions are the same as the ones used in Fig.~\ref{exp2}.}
	\label{exp1}
\end{figure*}

As shown in Appendix B, Eq.~(\ref{SF3}) has four roots: Two complex [Eq.~(\ref{roots1})] and two real [Eq.~(\ref{roots2})]. The two complex roots have no physical meaning, and therefore we turn our attention to the other remaining roots. Regarding the two real roots, only one of them is positive as time increases, while the other one is negative and thus corresponds to a suction process and not an injection. Therefore, we adopt the positive real root of Eq.~(\ref{SF3}) as the injection rate responsible to keep fixed the mode of largest growth rate $n_{\rm max}$ as the interface expands. Throughout this work, we'll refer to this pumping rate as $Q_{n_{\rm max}}(t)$.


It is worth noting that, by neglecting the wetting film contribution ($J_{0}=0$), Eq.~(\ref{SF3}) is significantly simplified, and one may readily find the solution as
\begin{eqnarray}
\label{Qnowet}
Q(t)=\frac{\pi^2 \beta (3n_{\rm max}^2-1)}{2{\rm Ca}_{g} (\beta-1) R(t)}.
\end{eqnarray}
Then, by using Eq.~(\ref{Revol}), one may rewrite the injection rate explicitly in terms of $t$ as $Q(t)\sim t^{-1/3}$ and thus reproducing the result previously obtained in Refs.~\cite{ShuwangPRL,mama1} in the absence of wetting effects.

Before analysing the efficiency of $Q_{n_{\rm max}}(t)$ in controlling the number of emerging fingers in the expanding interface, we first compare the injection rate $Q_{n_{\rm max}}(t)$, which takes into account the wetting effects ($J_{0}=3.8$), with the time-dependent flux given by Eq.~(\ref{Qnowet}), which neglects wetting ($J_{0}=0$). Figure \ref{Qcomp} illustrates this comparison by plotting the variation of these two different injection rates $Q(t)$ with respect to $R(t)$, for $n_{\rm max} = 5$, $\beta=10$, $q=25$, and ${\rm Ca}_{g} = 1000$.

It can be seen that in order to control the number of emerging fingers, the wetting case requires a slightly larger injection rate magnitude in comparison
with the non-wetting situation. In Ref.~\cite{pedrowet}, the authors demonstrate that, at the linear regime, wetting tends to stabilize the growth of interfacial instabilities by shifting $n_{\rm max}$ towards lower wave numbers and reducing the magnitude of the growth rate. Therefore, if one intends to keep the same mode $n_{\rm max}$ fixed in both situations (with and without wetting), a larger injection rate must be utilized when wetting effects are taken into account. The inset graph in Fig.~\ref{Qcomp} is plotted considering $\log_{10} Q(t)$ versus $\log_{10} R(t)$, and shows that, for this particular choice of parameters, wetting has a small impact in the dependence of $Q(t)$ with $R(t)$, i.e., in both cases $Q(t) \sim {R(t)}^{-1}$.

If one considers that the wetting film thickness scales as ${{\rm Ca}_l}^{2/3}$, which is defined by the fingertip velocity $\dot R = Q(t)/2\pi R(t)$, the film thickness would decrease in the radial direction since ${\dot R} \sim {R(t)}^{-2}$. In this scenario, the impact of the wetting film on the interfacial dynamics becomes less and less pronounced at later times, and this is the reason for both curves overlap for larger values of $R(t)$ in Fig.~\ref{Qcomp}.

To verify the efficiency of our linear-stability-based time-dependent injection rate $Q_{n_{\rm max}}(t)$ in controlling the interfacial development at fully nonlinear stages of the flow, when wetting effects are taking into account, we use our boundary integral formulation presented in Sec.~\ref{num} and Appendix A to describe the time evolution of the interface. Figure~\ref{exp2} plots the fully nonlinear interface evolution when the controlling pumping rate $Q_{n_{\rm max}}(t)$ is utilized to select a $5$-fold symmetry with $n_{\rm max} = 5$. In this figure, we set $\beta$ = 10, $q=25$, ${\rm Ca}_{g} = 1000$, and the initial condition is ${\cal R}(\theta,0)=1+0.02(\sin2\theta+\cos3\theta+\cos4\theta)$.

We initiate our discussion by surveying Fig.~\ref{exp2}(a), which illustrates a time overlaid plot of the evolving interface shown at equal time intervals. It is quite evident that the controlling injection rate $Q_{n_{\rm max}}(t)$ makes the interface evolve toward the targeted $5$-fold symmetric pattern. In addition, this fingering structure shows no tendency toward nonlinear ramification processes (such as finger-tip-splitting) and finger competition. It is a well-known fact that the main effects of the wetting film in radial Hele-Shaw flow are to restrain the development of both finger bifurcation and finger length variability~\cite{pedrowet}. However, wetting only delays the occurrence of these nonlinear pattern-forming mechanisms but does not fully suppress them~\cite{pedrowet,jackson}. Therefore, the absence of these mechanisms in the pattern depicted in Fig.~\ref{exp2}(a), even at such an advanced time regime, is caused by our time-dependent injection rate $Q_{n_{\rm max}}(t)$.

Note that this symmetric pattern evolves in time to the targeted $5$-fold structure set by our injection scheme, and, after that, the interface still expanding radially, but its shape is preserved, and the evolution becomes self-similar. Nevertheless, due to the very large final value of unperturbed radius ($R \approx 10^{15}$) utilized when plotting Fig.~\ref{exp2}(a), crucial information about the initial and intermediate regimes of the interfacial dynamics is lost. In particular, it seems that the pattern already initiates from a $5$-fold interface, and the final result is a consequence of a specific choice of the initial condition. However, that is not the case, as one may see by analyzing Figs.~\ref{exp2}(b) and~\ref{exp2}(c).

Figs.~\ref{exp2}(b) and~\ref{exp2}(c) provide supplementary details about how the system evolves from a nearly circular interface to the targeted $5$-fold self-similar final structure. Fig.~\ref{exp2}(b) illustrates the nonlinear variation of the rescaled perturbation amplitudes $|\zeta_{n} (t)|/R(t)$ ($2 \leq n \leq 15$) with $R(t)$ for the situation examined in Fig.~\ref{exp2}(a), while in Fig.~\ref{exp2}(c) we break the evolution depicted in Fig.~\ref{exp2}(a) into separate snapshots of the interface at different values of $R(t)$. To better visualize the interface morphology, each plot frame in Fig.~\ref{exp2}(c) is rescaled by its corresponding value of $R(t)$. Note that the absolute value $|\zeta_{n} (t)|/R(t) = \sqrt{{a^{2}_{n}}(t)+{b^{2}_{n}}(t)}/2R(t)$, where ${a}_{n}(t)$ and ${b}_{n}(t)$ are, respectively, the real valued cosine and sine Fourier amplitudes, is not obtained by utilizing the solution of the linear Eq.~(\ref{SF}), but rather is extracted directly from the fully nonlinear patterns generated by our numerical scheme. 

By analyzing Fig.~\ref{exp2}(b), one observes a rapid growth of modes 4 and 8, which dominate the dynamics from $R=1$ to $R=10^4$. This is in accordance with the snapshot of the interface for $R=2.2 \times 10^4$, where the pattern presents four wide fingers with flat tips, characterizing the onset of the classical tip-splitting phenomenon~\cite{pedrowet,jackson}. Based on this, one could expect that for subsequent times the interface would go through successive ramification processes ultimately leading to the formation of the usual branched patterns found in real experiments in radial Hele-Shaw cell~\cite{homsy1987viscous,casademunt2004viscous,mccloud1995experimental}. However, as $R$ increases, a different scenario is unveiled: The initially dominant modes 4 and 8 start to decay, and the system evolves into a complicated nonlinear stage characterized by intense interaction between the modes. In particular, note that mode 5, not present in the initial condition, is created and then selected by our controlling injection as the fastest growing mode. This influences the evolution of the interfacial pattern by changing its shape from a six-competing-fingers structure ($R=3.3 \times 10^6$) to an almost symmetric five-fingered interface ($R=7.2 \times 10^{10}$). Later, all the modes start to decay except by modes 5 and its harmonics (i.e., 10, 15, ...), which stabilize and remain constants as the interface expands, and thus reaching the self-similar stage of the dynamics ($R=2.1 \times 10^{17}$).  

We stress that the findings of Fig.~\ref{exp2} are also valid for other sets of parameters and initial conditions. In Fig.~\ref{exp1}, for example, other combinations of $n_{\rm max}$ and $\beta$ are utilized, and similar results to the ones already presented in Fig.~\ref{exp2} are obtained. Figure~\ref{exp1} plots the fully nonlinear interface evolution when the controlling pumping rate $Q_{n_{\rm max}}(t)$ is used [Fig.~\ref{exp1}(a)-Fig.~\ref{exp1}(d)], and the corresponding interface patterns for constant injection rate [Fig.~\ref{exp1}(e) and Fig.~\ref{exp1}(f)]. In Figs.~\ref{exp1}(a) and~\ref{exp1}(b) the time-dependent pumping rate we choose selects a $4$-fold symmetry with $n_{\rm max} = 4$. However, in Figs.~\ref{exp1}(c) and~\ref{exp1}(d) the time-dependent pumping rate intends to keep the number of fingers equals to five ($n_{\rm max} = 5$). While plotting these patterns, two values of $\beta$ have been used: $\beta$ = 10 (for the left column patterns) and $\beta$ = 100 (for the right column patterns). The interfaces are plotted in equal time intervals, and all the other physical parameters and initial conditions are the same as the ones used in Fig.~\ref{exp2}.

In Figs.~\ref{exp1}(a) and~\ref{exp1}(b) we see the establishment of four-fingered structures. At later times, the resulting patterns are still $4$-fold symmetric with no signs of nonlinear ramification processes and finger competition. In Figs.~\ref{exp1}(c) and~\ref{exp1}(d), it is also evident that the controlling injection rate $Q_{n_{\rm max}}(t)$ makes the interface evolve toward the targeted $5$-fold symmetric pattern, regardless of the value of $\beta$. 

This scenario is significantly changed when we consider the usual constant injection rate. As one can see by examining Fig.~\ref{exp1}(e), a constant injection rate leads to the development of a branched fingering pattern, that also exhibits variability among the lengths of the fingers. In addition, the number of fingers is not kept constant as the interface expands, making its growth disordered and unpredictable. In fact, the same behavior is identified in Fig.~\ref{exp1}(f) for $\beta=100$. Nevertheless, note that the larger value of viscosity ratio turns the pattern even more unstable, and nonlinear effects are enhanced. The morphologies shown in Figs.~\ref{exp1}(e) and~\ref{exp1}(f) are in agreement with interfacial patterns previously found in Refs.~\cite{pedrowet,jackson}. Observe that the growing $n$-fold patterns having a fixed number of fingers shown in Figs.~\ref{exp1}(a)-~\ref{exp1}(d) are dramatically different from the usual ramified shapes detected in Figs.~\ref{exp1}(e)-~\ref{exp1}(f) when a constant injection rate is employed. These findings confirm the efficiency of $Q_{n_{\rm max}}(t)$ in controlling the radial expansion of the interface in a Hele-Shaw cell in the presence of wetting film effects.

\begin{figure}[t]
	\centering
	\includegraphics[width=2.6 in]{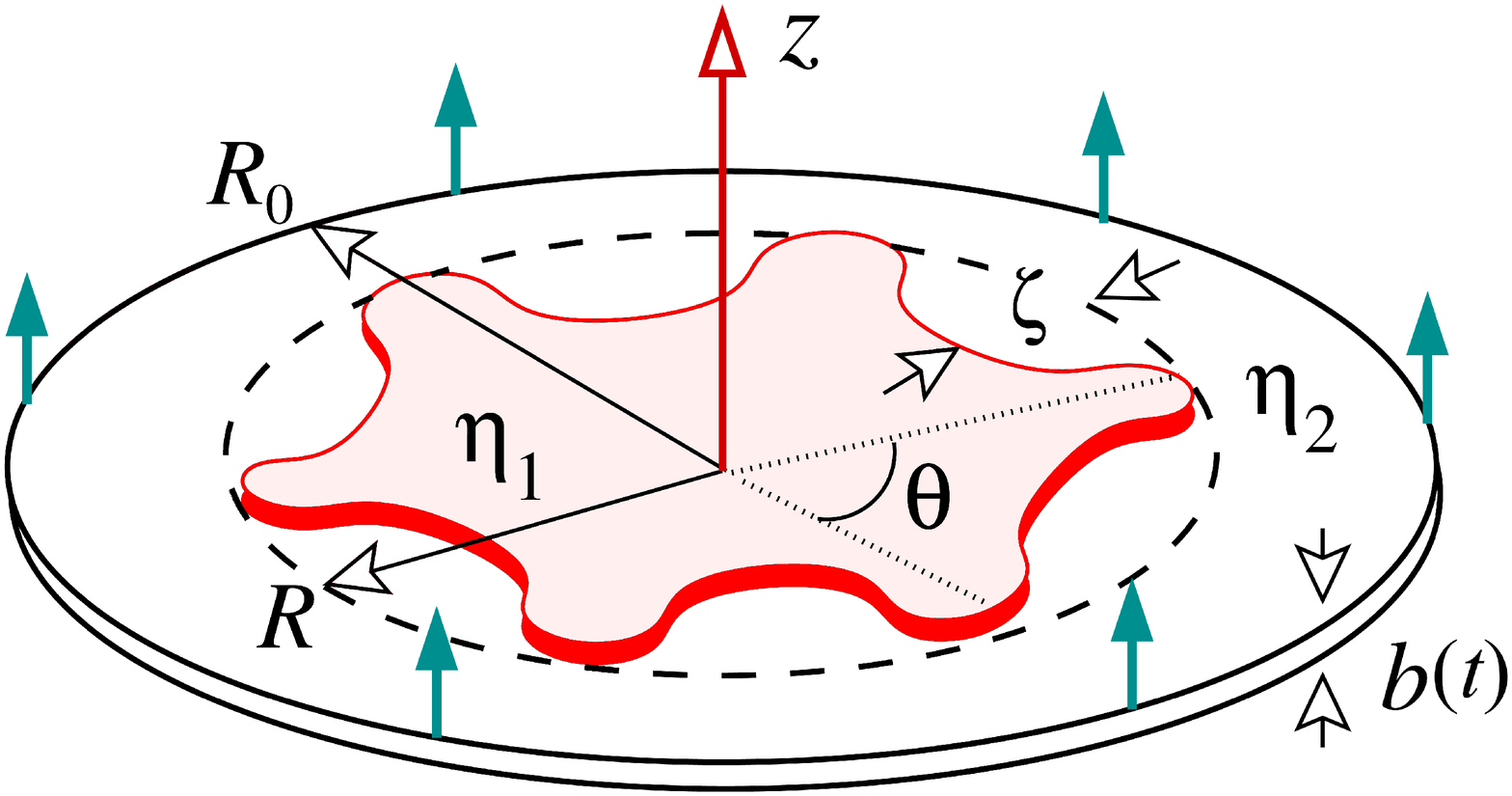}
	\caption{Representative sketch of the time-dependent gap flow in a radial Hele-Shaw cell.}
	\label{geom2}
\end{figure}

\section{TIME-DEPENDENT GAP HELE-SHAW FLOW}
\label{derivation2}

\subsection{Governing equations and linear growth rate}
\label{gov2}

The geometry of the time-dependent gap Hele-Shaw cell is sketched in Fig.~\ref{geom2}. Consider a Hele-Shaw cell of a variable gap width $b(t)$ containing two immiscible, incompressible, and viscous fluids. The upper plate of the cell can be lifted along the direction perpendicular to the cell plates ($z$-axis), and the initial fluid-fluid interface is circular, having radius $R_{0}$ and initial gap thickness $b_0=b(t=0)$. Here there is no injection. By using volume conservation, the time-dependent radius of the unperturbed
interface is given by
\begin{equation}
\label{R2}
R(t)=R_{0} \sqrt{ \frac{b_{0}}{b(t)}}.
\end{equation}

As in the case of the injection-driven flow, we use Darcy's law~(\ref{Darcy1}), the pressure jump condition~(\ref{pressure}), and the kinematic boundary condition~(\ref{kine}) to obtain a differential equation for the perturbation amplitudes. However, due to the lifting of the upper plate, the gap-averaged incompressibility condition~\cite{Shelley1} is now written as
\begin{equation}
\label{lift1}
{\bm \nabla}\cdot {\bf v}_{j}=-\frac{{\dot b}(t)}{b(t)},
\end{equation}
where ${\dot b}(t)=db/dt$ is the upper plate velocity along the $z$-axis. Moreover, the velocity potential obeys a Poisson equation
\begin{equation}
\label{lift2}
\nabla^2\phi_{j}=\frac{{\dot b}(t)}{b(t)}.
\end{equation}
Another important difference between the injection-driven flow presented in Sec.~\ref{gov} and the time-dependent gap setup is the fact that here is the nonwetting fluid 2 that displaces the wetting fluid 1, so that $\alpha_{c}=\pi$ and ${\rm Ca}_{l}=\eta_{1} V/\sigma$ in the pressure jump condition~(\ref{pressure}).

By following the standard steps described in the injection-driven situation of Sec.~\ref{gov}, a dimensionless equation of the form given by Eq.~(\ref{result}) is obtained for the lifting Hele-Shaw case, where now
\begin{eqnarray}
\label{growth2}
\lambda(n) &=& \frac{1}{1 + w(n)} \Bigg [- \frac{\dot b}{2 b} \left(1 + \frac{\beta-1}{\beta+1} |n|\right) \nonumber\\
&&- \frac{\pi b^{2}}{4 {\rm Ca}_{g} (qR)^{3} (\beta+1)} |n| (n^{2} - 1) \Bigg ],
\end{eqnarray}
is the linear growth rate, and
\begin{equation}
\label{w2}
w(n)= |n| J_{0}  \frac{b}{9qR(\beta+1)} \left ( \frac{24 b}{\dot b {\rm Ca}_{g} q R} \right )^{1/3}
\end{equation}
is related to the wetting film contribution. We have nondimensionalized Eqs.~(\ref{growth2}) and~(\ref{w2}) as follows: (i) in-plane lengths are rescaled by $R_{0}$; (ii) $b(t)$ is scaled on its initial value $b_{0}$; (iii) likewise, time is rescaled by the 
characteristic time $T=b_{0} / |{\dot b}_{0}|$. The global capillary number is
\begin{equation}
\label{cap2}
{\rm Ca}_{g} = \frac{12 \eta_{1} {\dot b}_{0}  }{\sigma}
\end{equation}
and $q = R_{0}/b_{0}$ is the initial aspect ratio. 

The expressions~(\ref{growth2}) and (\ref{w2}) represent the linear equations of the viscous fingering problem in a lifting Hele-Shaw cell, taking into consideration the contributions from wetting film effects. Note that the situation in which wetting effects are neglected can be readily obtained by setting $J_{0}=0$. In this case, we do recover the linear growth rate derived 
in the literature in the absence of wetting effects~\cite{Shelley1,Ben3}.

\subsection{Numerical scheme}
\label{num2}

For the shrinking interface, it's more convenient to work with the pressure field instead of velocity potential~\cite{Zhao20,Zhao18}. Therefore, we define the modified pressure $\displaystyle \tilde{p}(\mathbf{x})=p(\mathbf{x})- \frac{\dot{b}(t)}{4b^3(t)}|\textbf{x}|^2$, which is harmonic and satisfies a double layer potential
\[\tilde{p}(\textbf{x})=\int_{\Gamma}{\gamma(\mathbf{x}')}\left(\frac{\partial \ln|\textbf{x}-\textbf{x}'|}{\partial\textbf{n}(\textbf{x}')}+1\right)ds(\textbf{x}').\]
Assuming that the outer fluid 2 is air (i.e., $\beta=0$), we only need to solve the inner fluid 1 problem. Using the dimensionless boundary conditions of the problem, we show that ${\gamma}$ satisfies a Fredholm integral equation of the second kind
\begin{eqnarray}
&{\gamma(\mathbf{x})}&+\frac{1}{\pi}\int_{\Gamma}{\gamma(\mathbf{x}')}\left(\frac{\partial \ln|\textbf{x}-\textbf{x}'|}{\partial \textbf{n}(\textbf{x}')}+1\right)ds(\textbf{x}') \nonumber\\
&=&\frac{1}{{q^3\rm Ca}_{g}}\left[-\frac{2q}{b(t)}(1+J_0|{\rm Ca}_l|^{2/3})+\frac{\pi}{4}\kappa\right] \nonumber\\
&&-\frac{\dot{b}(t)}{2b^3(t)}|\textbf{x}|^2,\label{integrolp}
\end{eqnarray}
and once we obtain ${\gamma(\mathbf{x})}$, $\tilde{V}(\mathbf{x})$ can be computed via Dirichlet-Neumann map~\cite{LapMCD}
\begin{equation}
\tilde{V}(\mathbf{x})=-\frac{b^2(t)}{2\pi}\int_{\Gamma}{\gamma}_{s}(\mathbf{x}')\frac{(\mathbf{x}'-\mathbf{x})^\perp\cdot \mathbf{n}(\mathbf{x})}{|\mathbf{x}'-\mathbf{x}|^2}ds(\textbf{x}').
\label{tV}
\end{equation}
Then, we write the normal velocity of the interface as
\begin{equation}
V(\mathbf{x})=-\frac{b^2(t)}{2\pi}\int_{\Gamma}{\gamma}_{s}(\mathbf{x}')\frac{(\mathbf{x}'-\mathbf{x})^\perp\cdot \mathbf{n}(\mathbf{x})}{|\mathbf{x}'-\mathbf{x}|^2}ds(\textbf{x}')-\frac{\dot{b}(t)}{2b(t)}\textbf{x}\cdot\textbf{n}\label{V}.
\end{equation}
Eq.~(\ref{integrolp}) is well-conditioned and coupled with Eq.~(\ref{V}) via $\displaystyle {\rm Ca}_l=\frac{\mu_1R_0}{\sigma T}V$. This system can be solved efficiently using an iterative method such as GMRES~\cite{GMRES}, and $V$ is computed using the Picard iteration described in Section~\ref{num}. Once the normal velocity $V$ is determined, the interface is evolved by utilizing Eq.~(\ref{nonexp}). Similar to the expanding case presented in Section~\ref{num}, here, the long-time simulations are also expensive, and in Appendix C we introduce a rescaling idea designed for the lifting Hele-Shaw flow to improve the efficiency of the method. 

\subsection{Shrinking evolution with fixed number of fingers}
\label{self2}

Similar to what we have done in Sec.~\ref{self}, here our goal is to develop a specific time-dependent lifting speed ${\dot b}(t)$ expression that keeps $n_{\rm max}$ fixed as the interface shrinks radially to the center of the cell. For the lifting Hele-Shaw setup, the unperturbed radius $R(t)$ satisfies
\begin{equation}
\label{Revol2}
\dot R = - \frac{{\dot b}(t)}{2 R(t) {b^2(t)}},
\end{equation}
and the linear evolution of the rescaled amplitudes is given by Eq.~(\ref{SF}), with
\begin{eqnarray}
\label{modlam2}
\Lambda(n)=\lambda(n) + \frac{\dot b}{2 (Rb)^2}.
\end{eqnarray}
By setting the derivative with respect to $n$ of Eq.~(\ref{modlam2}) equal to zero, utilizing Eqs.~(\ref{growth2}) and~(\ref{w2}), and using the fact that $R(t)=1/\sqrt{b(t)}$, we obtain
\begin{eqnarray}
\label{SF4}
&-&\left(\frac{\beta-1}{\beta+1} \right)\frac{\dot b}{2b}  \nonumber\\
&+& \frac{J_{0}}{18q(\beta+1)} \left( \frac{24}{q  {\rm Ca}_{g}}\right)^{1/3} {\dot b}^{2/3} b -\frac{\pi (3n_{\rm max}^2-1)}{4q^3 {\rm Ca}_{g} (\beta+1)} b^{7/2} \nonumber\\
&-& \frac{\pi J_{0} n_{\rm max}^3}{18 (\beta+1)^2 q^4 {\rm Ca}_{g}} \left(\frac{24}{q {\rm Ca}_{g}} \right)^{1/3} {\dot b}^{-1/3} b^{11/2}=0.
\end{eqnarray}
Eq.~(\ref{SF4}) can also be identified as a quartic equation after one defines $x \equiv {{\dot b}}^{1/3}$. Therefore, all the discussion related to the solutions of the quartic Eq.~(\ref{SF3}) and the appropriate way to deal with this kind of expression, as well as the information provided by Appendix B, remain valid for Eq.~(\ref{SF4}). Hence, the positive real solution of Eq.~(\ref{SF4}), i.e., ${\dot b}_{n_{\rm max}}(t)$, is the adequate time-dependent gap speed needed to maintain the number of fingers fixed in a lifting Hele-Shaw cell flow when the displaced fluid is a wetting fluid. On the other hand, the negative real solution corresponds to a squeezing process and can be disregarded.

\begin{figure}[t]

	\includegraphics[scale=0.55]{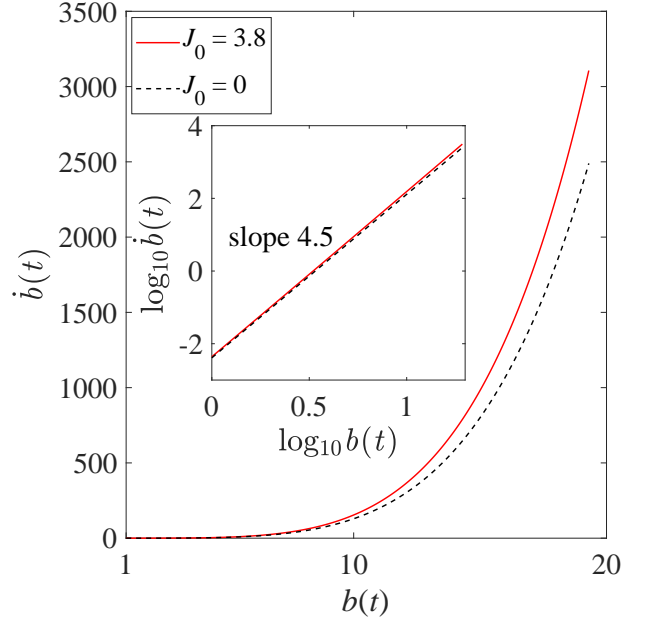}
	
	\caption{Time-dependent lifting speed ${\dot b}(t)$ as a function of $b(t)$, for the cases with ($J_{0}=3.8$), and without wetting ($J_{0}=0$). The solid curve corresponds to the gap speed ${\dot b}_{n_{\rm max}}(t)$, while the dashed one is obtained by considering Eq.~(\ref{bnowet}). Here $n_{\rm max} = 3$, $\beta=0$, $q=112$, and ${\rm Ca}_{g} = 7.1 \times 10^{-3}$.}
	\label{bcomp}
\end{figure}

\begin{figure*}[t]
	\includegraphics[scale=0.509]{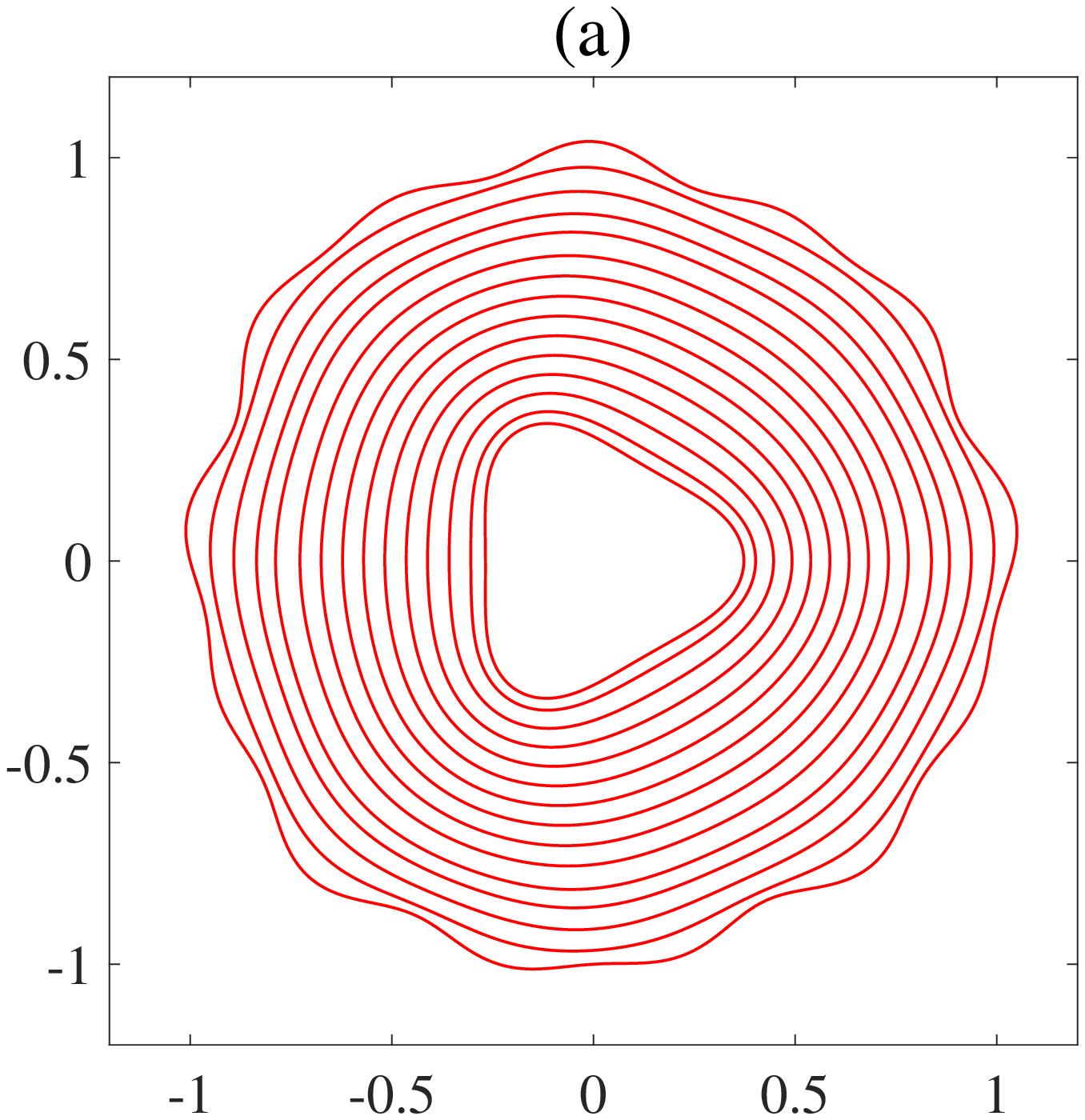}
	\includegraphics[scale=0.5]{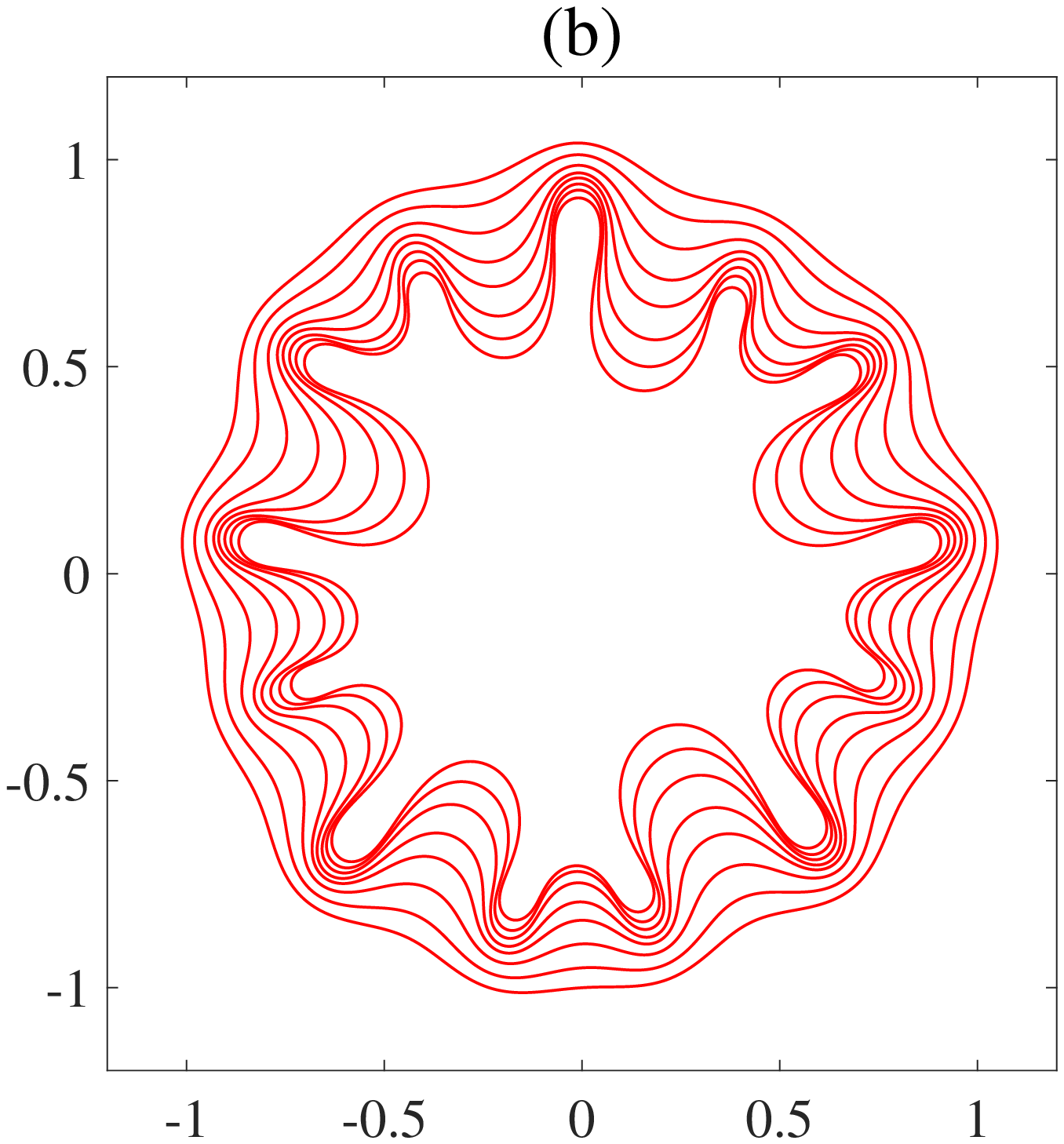}\\
	\includegraphics[scale=0.5]{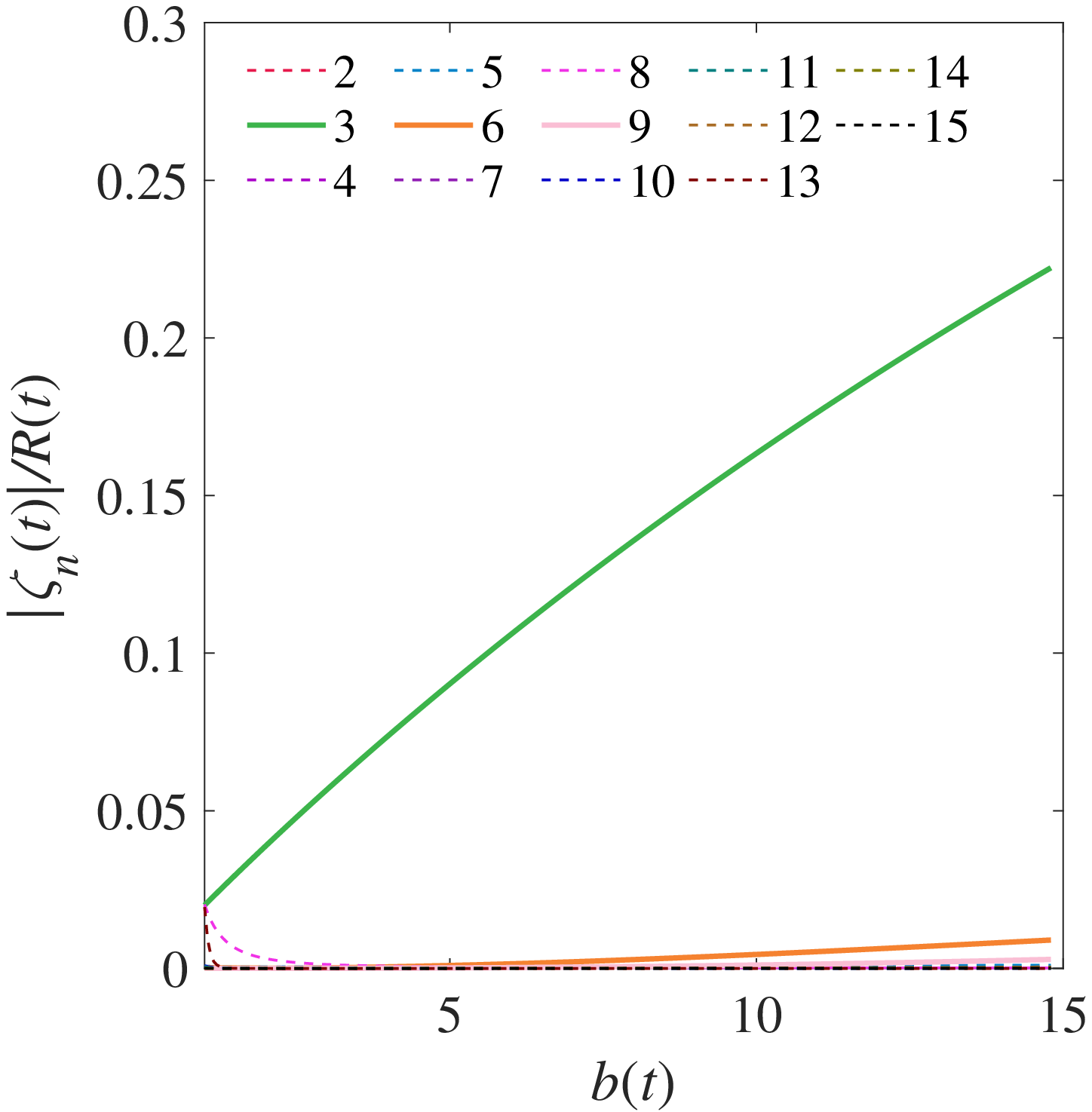}
	\includegraphics[scale=0.5]{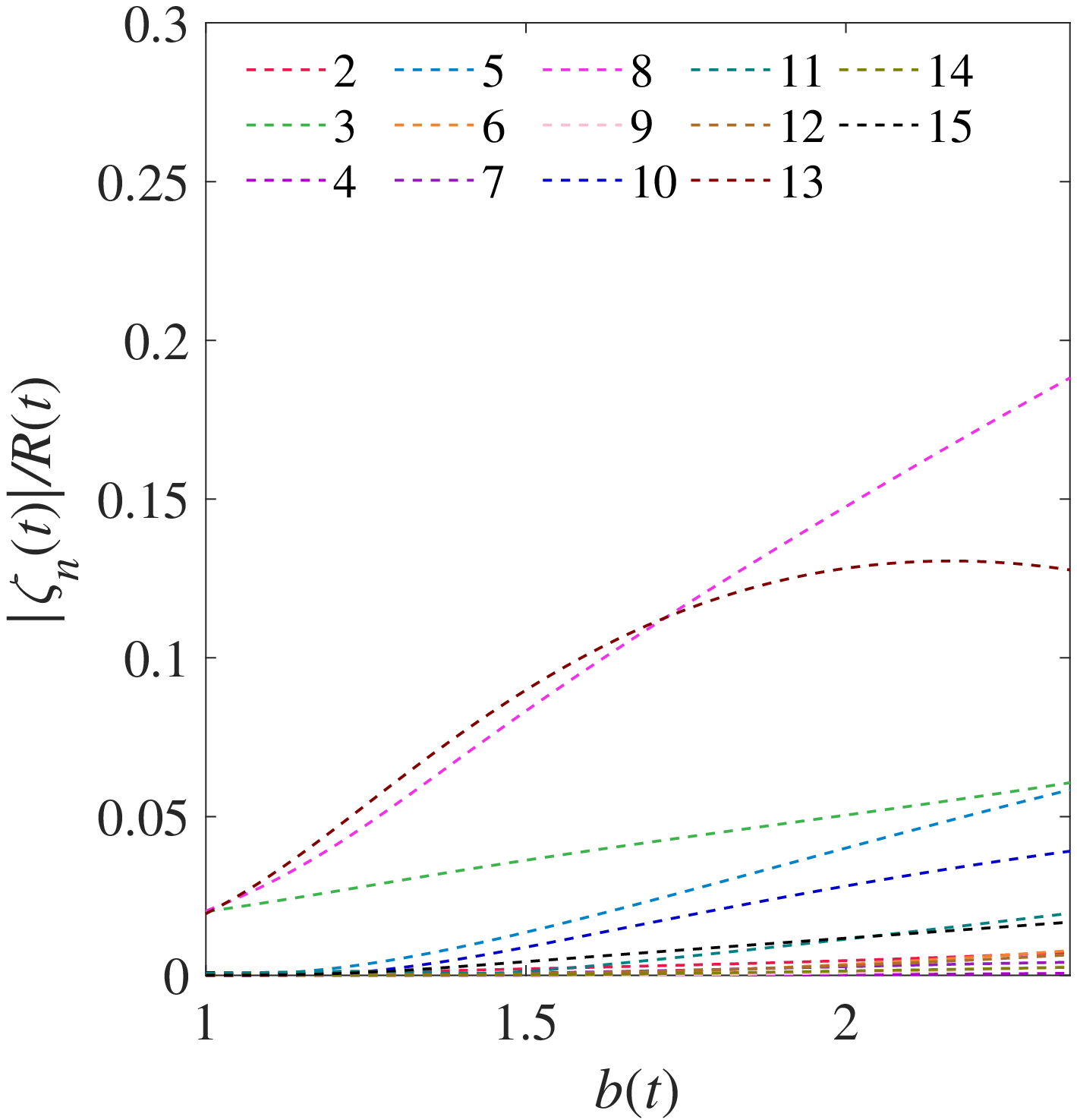}

	\caption{Time evolution of the fully nonlinear interfacial patterns produced by (a) the time-dependent gap speed ${\dot b}_{n_{\rm max}}(t)$ obtained by solving Eq.~(\ref{SF4}), and by (b) the usual constant gap speed ${\dot b}=1$. The corresponding variation of the rescaled perturbation amplitudes $|\zeta_{n} (t)|/R(t)$ with $b(t)$, for all the modes in the interval $2 \leq n \leq 15$, is depicted in the bottom panels. In (a) we impose that mode $n_{\rm max} = 3$ is the fastest growing. Here $\beta=0$, $q=112$, ${\rm Ca}_{g} = 7.1 \times 10^{-3}$, and the initial condition is ${\cal R}(\theta,0)=1+0.02(\cos3\theta+\cos8\theta+\sin13\theta)$.}
	\label{exp3}
\end{figure*}


Note that by neglecting the wetting film contribution ($J_{0}=0$), Eq.~(\ref{SF4}) is reduced to 
\begin{eqnarray}
\label{bnowet}
{\dot b}(t) = -\frac{\pi (3n_{\rm max}^2-1)}{2q^3 {\rm Ca}_{g} (\beta-1)} {b(t)}^{9/2}.
\end{eqnarray}
This differential equation can be easily solved to find $b(t) \sim t^{-2/7}$. Within this limit, our results agree with previous works~\cite{Zhao18,mama1} that analyzed the controlling problem in the absence of wetting effects for the lifting Hele-Shaw setup.

Figure~\ref{bcomp} compares the lifting gap speed ${\dot b}_{n_{\rm max}}(t)$ designed to account for wetting effects ($J_{0}=3.8$), with the non-wetting ($J_{0}=0$) gap speed given by Eq.~(\ref{bnowet}). Both speeds ${\dot b}(t)$ are plotted as a function of $b(t)$, for $n_{\rm max} = 3$, $\beta=0$, $q=112$, and ${\rm Ca}_{g} = 7.1 \times 10^{-3}$.

First, note that to keep the same mode $n_{\rm max}$ fixed in both cases, one needs to utilize a larger gap speed in the wetting system when compared to the non-wetting situation. This is due to the stabilizing effect provided by the wetting film. Also, we observe by analysing the inset plot that the dependence of ${\dot b}(t)$ with $b(t)$ is nearly the same for both situations, i.e., ${\dot b}(t) \sim {b(t)}^{9/2}$. These conclusions are in line with what has been found previously for the injection-driven flow presented in Sec.~\ref{self}. On the other hand, note that the distance between the two curves increases as $b(t)$ gets larger. In the shrinking evolution, the fingertip velocity is given by ${\dot R} = - {\dot b} (t)/2R(t){b(t)}^2 =  - {\dot b} (t)/2{b(t)}^{3/2}$, and since ${\dot b}(t) \sim {b(t)}^{9/2}$, one concludes that ${\dot R} \sim {b(t)}^{3}$. Therefore, in opposition to the expanding case, in the shrinking situation, the film thickness increases as the interface move radially towards the center of the Hele-Shaw cell, enhancing wetting effects at later times.

We close this section by discussing Fig.~\ref{exp3}, which illustrates the fully nonlinear fluid-fluid interface evolution obtained by utilizing (a) the time-dependent gap speed ${\dot b}_{n_{\rm max}}(t)$ with $n_{\rm max} = 3$, and (b) the usual constant lifting speed ${\dot b}=1$. In the bottom panels, we plot the corresponding variation of $|\zeta_{n} (t)|/R(t)$ with $b(t)$ for all the modes in the interval $2 \leq n \leq 15$. Here, we consider that the outer fluid 2 is air while the inner fluid 1 is a very viscous oil and hence $\beta=0$. In addition, $q=112$, ${\rm Ca}_{g} = 7.1 \times 10^{-3}$, and the initial condition is ${\cal R}(\theta,0)=1+0.02(\cos3\theta+\cos8\theta+\sin13\theta)$. 

By examining Fig.~\ref{exp3}(a), we observe an initial slightly perturbed interface evolving to a well-behaved morphology as the interface shrinks. The final pattern is not only 3-fold symmetric, but also does not exhibit signs of finger competition. These visual conclusions are in agreement with the corresponding evolution of the rescaled perturbation amplitudes $|\zeta_{n} (t)|/R(t)$ depicted in the bottom panel, where we can verify that mode 3 is selected as the fastest growing mode by our lifting speed scheme. As a matter of fact, ${\dot b}_{n_{\rm max}}(t)$ also promotes the enhanced growth of the harmonic modes 6, 9, and 12, over all the other remaining modes. That effect was also detected in Fig.~\ref{exp2} for the expanding interface case. Note that in the injection-drive situation, the interface expands radially over the very large domain of fluid 2, and self-similarity is achieved for large values of $R$. On the other hand, in the time-dependent gap flow, the interface shrinks radially over the finite domain of fluid 1, and for the values of $R$ used here, we did not observe the establishment of a self-similar evolution.

Now we compare the interfacial evolution depicted in Fig.~\ref{exp3}(a) with the case illustrated in Fig.~\ref{exp3}(b), which considers the lifting Hele-Shaw flow subjected to a constant lifting speed ${\dot b}=1$, so that $b(t)=1+t$. We choose this specific situation due to the fact that it is the most common case investigated in experimental and theoretical studies in lifting Hele-Shaw flows~\cite{Roy,Anke,Ben3,Anke2,Nase,Pedro,lift1,lift2}. Inspection of Fig.~\ref{exp3}(b) shows that the time evolution under ${\dot b}=1$ is considerably different from the equivalent situation utilizing ${\dot b}_{n_{\rm max}}(t)$ [Fig.~\ref{exp3}(a)]. First, we note the formation of 13 inward competing fingers with larger amplitudes than the fingers formed in the pattern depicted in Fig.~\ref{exp3}(a). Moreover, these wide inward-pointing fingers are alternated by the sharp, outward-pointing fingers of the inner fluid. These morphological features are in line with the findings of Ref.~\cite{Pedro}, which analyzed the effects of wetting in the lifting Hele-Shaw cell problem at the weakly nonlinear regime. One can also observe that the evolution of $|\zeta_{n} (t)|/R(t)$ shows a tendency towards the growth of all the modes as the interface shrinks, without selection of a specific mode. This impacts the interfacial evolution by turning it more unstable and disordered when compared to the controlled flow promoted by ${\dot b}_{n_{\rm max}}(t)$ in Fig.~\ref{exp3}(a).

\section{Conclusion}
\label{conclude}

Injection-driven flow in radial Hele-Shaw cells results in ramified patterns if the injection rate is constant in time. The emerging structures are characterized by the occurrence of finger ramification and finger competition events~\cite{homsy1987viscous,casademunt2004viscous,mccloud1995experimental}. Likewise, time-dependent gap flow in lifting Hele-Shaw cells leads to complex pattern morphologies~\cite{Ben2,Roche,Shelley1,Roy,Anke,Tarafdar,Ben3,Anke2,Tarafdar2,Nase,Diasmaster,Stone} if the cell's gap width grows linearly
with time, i.e., lifting gap speed is constant in time. In this case, the resulting shapes are formed due to the intense competition among the fingered structures.

In many practical applications, the emergence of these hydrodynamic instabilities is undesirable and because of that much attention has been devoted to devising strategies for controlling the growth of such patterns. However, all existing studies that somehow try to control the emergence of these interfacial disturbances neglect the effects of the wetting film left behind by the displaced fluid during the flow, even though a considerable number of works~\cite{Park,Tab,Schwartz,Saf2,Rei1,Rei2,Max,Russo,Max_amp,pedrowet,jackson,Pedro,pedroadh} have pointed to the fact that wetting has an important role in the nonlinear finger formation process.

Motivated by these facts, in this work, we have explored the possibility of controlling
the development of interfacial instabilities when wetting effects are taken into account. This was done by properly manipulating the injection rate $Q(t)$ (for expanding case) and the gap lifting speed ${\dot b}(t)$ (for shrinking situation), with the ultimate goal of designing feasible and more accurate controlling strategies that could be utilized for technological and industrial purposes. 

First, by taking into account wetting film effects, we employed a linear stability analysis for obtaining the optimal injection and lifting protocols, i.e., the strategies intended to control the total number of resulting fingers arising at the fluid-fluid interface for the injection-driven and lifting flows, respectively.
The consideration of the wetting film led to an increase in the magnitude of $Q(t)$ and ${\dot b}(t)$ in comparison to the non-wetting strategies. Then, we utilized a fully nonlinear boundary integral scheme to verify the effectiveness of these linear-stability-based controlling strategies in the advanced time regime of the dynamics.

Our numerical results show that these time-dependent protocols are indeed capable of promoting controlled development of the interface even at fully nonlinear stages of the flow. In particular, for injection-driven flow subjected to our time-dependent scheme, the resulting patterns are $n$-fold symmetric structures for which the number of fingers is kept constant as the interface grows radially, and no signs of nonlinear finger ramification and finger competition were identified. Moreover, the expanding interface evolved self-similarly after reaching a certain size. On the other hand, we have identified that our time-dependent lifting speed protocol is also successful in prescribing the number of emerging fingers during the shrinking evolution in the lifting flow. In this last situation, finger competition was suppressed, however, self-similar evolution was not achieved.  

\begin{acknowledgments}
	
S. L. acknowledges the support from the National Science Foundation, Division of Mathematical Sciences grant DMS-1720420. J. L. acknowledges partial support from the NSF through grants DMS-1714973, DMS-1719960, DMS-1763272, and the Simons Foundation (594598QN) for a NSF-Simons Center for Multiscale Cell Fate Research. J. L. also thanks the National Institutes of Health for partial support through grants 1U54CA217378-01A1 for a National Center in Cancer Systems Biology at UC Irvine and P30CA062203 for the Chao Family Comprehensive Cancer Center at UC Irvine.
W. B. acknowledges the support from the Academic Research Fund of the Ministry of Education of Singapore grant No. MOE2019-T2-1-063 (R-146-000-296-112).
Part of the work was done when the last two authors (W. B. and S. L.) were 
visiting the Institute of Mathematical Science at the National University of 
Singapore in 2020.
\end{acknowledgments}

\appendix
\section{Rescaling scheme for injection-driven flow}

We introduce a new rescaled space and time frame $(\bar{\bf x}, \bar t)$ such that
\begin{equation}
\mathbf{x}=\bar R(\bar t)\mathbf{\bar x}(\bar t,\alpha)
\label{xs}
\end{equation}
and
\begin{equation}
\bar t=\int_{0}^t \frac{1}{\rho(t^\prime)}dt^\prime, 
\label{ts}
\end{equation} 
where $\bar R(\bar t)$ is the space scaling factor representing the size of the interface, and  $\mathbf{\bar x}$ is the position vector of the scaled interface with  parametrization $\alpha$. The time scale function $\rho(t)$ maps the original time $t$ to the new time $\bar t$. In general, $\rho(t(\bar t))={\bar \rho}(\bar t)$ can be chosen arbitrarily to make the interface evolve in the new frame at any speed. The rescaled normal velocity $\bar{V}$  satisfies
\begin{equation}
\bar{V}(\bar t)=\frac{\bar{\rho}}{\bar R}V(t(\bar t))-\frac{\mathbf{\bar x}\cdot \mathbf{n}}{\bar R}\frac{d\bar R}{d\bar t},
\label{Vrelation}
\end{equation}
where $V$ is the original normal velocity. In the rescaled frame, we require that the area enclosed by the interface remains constant $\bar{A}(\bar t)=\bar{A}(0)$. That is, the integration of the normal velocity along the interface in the scaled frame vanishes $\displaystyle \int_{\bar{\Gamma}(\bar t)}\bar{V}d\bar{s}=0$.
As a consequence, 
\begin{equation}
\frac{d\bar R}{d\bar t}=\frac{\bar{\rho}\bar{Q}}{2\bar{A}(0)\bar R}.
\label{scale}
\end{equation}
Then, by taking $\displaystyle \rho(\bar{t})=\frac{2\bar A \bar R^2(\bar t)}{ \bar Q}$ we make $\bar{R}(\bar t)$ evolves exponentially fast in the rescaled frame 
\begin{equation}
\bar{R}(\bar{t})=\exp(\bar t). 
\label{newR}
\end{equation}
Taking $\bar \gamma(\mathbf{\bar x})=\gamma(\mathbf{x})\bar R(\bar t)$, we next rewrite the integral equation (\ref{eqmu}) in the rescaled frame as
\begin{eqnarray}
&&~~\frac{1}{2}\left(\frac 1 \beta+1\right)\bar \gamma(\mathbf{\bar x}) \nonumber\\
&+&\frac{1}{2\pi}\left(\frac 1\beta-1\right)\int_{\bar{\Gamma}(\bar t)}\bar \gamma(\mathbf{\bar x}')\left[\frac{\partial \ln|\mathbf{\bar x}-\mathbf{\bar x}'|}{\partial \mathbf{n}(\mathbf{\bar x}')}+\bar R(\bar t)\right]d\bar{s}(\mathbf{\bar x}') \nonumber\\
&=&-\frac{1}{{\rm Ca}_{g}}\left[\bar R(\bar t)2q(1+J_0|{\rm Ca}_l|^{2/3})+\frac{\pi}{4}\bar\kappa\right] \nonumber\\
&&-\left(\frac 1\beta-1\right)\bar R(\bar t) \frac{\bar{Q}}{2\pi}(\ln(\bar R(\bar t))+\ln|\mathbf{\bar x}|).
\label{mub}
\end{eqnarray}
Similarly, we compute  the normal velocity in the rescaled frame as
\begin{eqnarray}
\label{Vb}
\bar V(\mathbf{\bar x})&=&\frac{\bar{A}}{\pi\bar{Q}}\bigg(\frac{1}{\bar R}\int_{\bar{\Gamma}(\bar t)}\bar \gamma_{\bar{s}}(\mathbf{\bar x}')\frac{(\mathbf{\bar x}'-\mathbf{\bar x})^{\perp}\cdot\mathbf{\bar{n}}(\bar{\mathbf{x}})}{|\mathbf{\bar x}'-\mathbf{\bar x}|^2}d\bar{s}(\mathbf{\bar x}')\nonumber\\
&&+ \bar{Q}\frac{\mathbf{\bar x}\cdot \mathbf{\bar{n}}}{|\mathbf{\bar x}|^2}\bigg)-\mathbf{\bar x}\cdot \mathbf{\bar{n}},
\end{eqnarray}
where $\mathbf{\bar x}^{\perp}=(\bar{x}_2,-\bar x_1)$. The interface is evolved in the scaled frame through
\begin{equation}
\frac{d\bar{\textbf{x}}(\bar t,\alpha)}{d\bar t}\cdot \textbf{n}=\bar V(\bar t,\alpha).
\label{evolvesc}
\end{equation}

\section{Solutions of Eqs.~(\ref{SF3}) and~(\ref{SF4})}

In order to show the solutions of Eq.~(\ref{SF3}), we first rewrite it as
\begin{equation}
\label{poli}
Ax^4+Bx^3+Cx+D=0,
\end{equation}
where
\begin{eqnarray*}
\label{ABCD}
A&=&\frac{1}{2\pi R^2}\left(\frac{\beta -1}{\beta+1} \right), \nonumber\\
B&=& \frac{J_{0}\beta}{18\pi (\beta+1) q R^3} \left(\frac{24\pi R q^2}{{\rm Ca}_{g}} \right)^{1/3}, \nonumber\\
C&=&-\frac{\pi \beta (3n_{\rm max}^2-1)}{4{\rm Ca}_{g} (\beta+1) R^3},  \nonumber\\
D&=&-\frac{\pi J_{0} \beta^2 n_{\rm max}^3}{18 (\beta+1)^2 q R^4 {\rm Ca}_{g}} \left(\frac{24\pi R q^2}{{\rm Ca}_{g}} \right)^{1/3},
\end{eqnarray*}
and $x=Q^{1/3}$. The narute of the roots of the quartic Eq.~(\ref{poli}) is mainly determined by the sign of its discriminant~\cite{root1}
\begin{eqnarray}
\label{discri}
\Delta&=&256A^3D^3-192A^2BCD^2-27A^2C^4 \nonumber\\
&&-6AB^2C^2D-27B^4D^2-4B^3C^3,
\end{eqnarray}
which is negative and therefore indicates the existence of two distinct real roots and two complex conjugate non-real roots. Following Ferrari's method~\cite{root2}, the two complex roots can be written as
\begin{equation}
\label{roots1}
x_{1,2}=-\frac{B}{4A}-S\pm \frac{1}{2}i\sqrt{\bigg\rvert -4S^2-2p+\frac{q}{S}\bigg\rvert},
\end{equation}
and the two real roots are
\begin{equation}
\label{roots2}
x_{3,4}=-\frac{B}{4A}+S\pm \frac{1}{2}\sqrt{-4S^2-2p-\frac{q}{S}},
\end{equation}
where
\begin{eqnarray*}
\label{pqs}
i^2&=&-1,~~p=-\frac{3B^2}{8A^2}, ~~q=\frac{B^3+8A^2C}{8A^3}<0,\nonumber\\
S&=&\frac{1}{2}\sqrt{-\frac{2}{3}p+\frac{1}{3A}\left(X+\frac{12AD-3BC}{X}\right)},\nonumber\\
~~~~~~~~X&=&\sqrt[3]{\frac{27(B^2D+AC^2)+\sqrt{-27\Delta}}{2}}.
\end{eqnarray*}

By the reasons mentioned in Sec.~\ref{self}, we adopt the cubed positive real root ${{x}_{3}^{3}}$ as our injection strategy $Q_{n_{\rm max}}(t)$.

Regarding Eq.~(\ref{SF4}), after setting $x={\dot b}^{1/3}$, it acquires exactly the same format of Eq.~(\ref{poli}), but with coefficients given by
\begin{eqnarray*}
	\label{ABCD2}
	A&=&-\left(\frac{\beta-1}{\beta+1} \right)\frac{1}{2b}, \nonumber\\
	B&=& \frac{J_{0}}{18q(\beta+1)} \left( \frac{24}{q  {\rm Ca}_{g}}\right)^{1/3} b, \nonumber\\
    C&=&-\frac{\pi (3n_{\rm max}^2-1)}{4q^3 {\rm Ca}_{g} (\beta+1)} b^{7/2},\nonumber\\
	D&=&- \frac{\pi J_{0} n_{\rm max}^3}{18 (\beta+1)^2 q^4 {\rm Ca}_{g}} \left(\frac{24}{q {\rm Ca}_{g}} \right)^{1/3} b^{11/2}.
\end{eqnarray*}
Therefore, the expressions~(\ref{roots1}) and~(\ref{roots2}) are also the solutions for the lifting case presented in Sec.~\ref{self2} and we utilize ${{x}_{3}^{3}}$ as our lifting speed ${\dot b}_{n_{\rm max}}(t)$.

\section{Rescaling scheme for time-dependent gap flow}

Similar to what has been done in Appendix A, here we also introduce the new rescaled space and time frame $(\bar{\bf x}, \bar t)$, and Eqs.~(\ref{xs}),~(\ref{ts}), and~(\ref{Vrelation}) remain valid for the time-dependent gap flow. 

Using volume conservation for the inner viscous fluid 1, we have 
\begin{equation}
\bar R^{-1}\frac{d\bar R}{d\bar t}=-\frac{\bar{\rho}\dot{b}(t(\bar t))}{2b(t(\bar t))},\label{frame}
\end{equation}
where dot means the time derivative in the original frame. Substituting Eq. (\ref{frame}) into Eq. (\ref{Vrelation}), and using Eq. (\ref{tV}), we have that $\displaystyle \bar V= \frac{\bar{\rho}}{\bar R}\tilde{V}(t)$. Here we assume $\displaystyle \frac{\dot{b}(t(\bar t))}{2b(t(\bar t))}=f(\bar R)$ and we choose $\displaystyle \bar R(\bar t)=(1+\bar t)^{-1/4}$. Thus, we obtain $\displaystyle \bar \rho=\frac{{\bar R}^4}{2f(\bar R)}$.

Taking $\displaystyle {\gamma}=\bar{\gamma}{\bar R}^{-1}$, Eq.~(\ref{integrolp}) can be rewritten as
\begin{eqnarray}
&\bar{\gamma}&(\mathbf{\bar x})+\frac{1}{\pi}\int_{\bar{\Gamma}(\bar t)}\bar{\gamma}(\mathbf{\bar x}')\left[\frac{\partial \ln|\mathbf{\bar x}-\mathbf{\bar x}'|}{\partial \textbf{n}(\mathbf{\bar x}')}+\bar R(\bar t)\right]ds(\mathbf{\bar x}') \nonumber\\
&=& \frac{1}{{q^3\rm Ca}_{g}}\left[-\frac{2q\bar R}{b(\bar t)}(1+J_0|{\rm Ca}_l|^{2/3})+\frac{\pi}{4}\kappa\right] \nonumber\\
&&-\frac{\dot{b}(\bar t){\bar R}^3}{2b^3(\bar t)}|\mathbf{\bar x}|^2,\label{integrolpres}
\end{eqnarray}
and we compute the normal velocity in the rescaled frame by utilizing
\begin{equation}
\bar V(\bar{\textbf{x}})=-\frac{b^2(\bar t)\bar{\rho}}{2\pi \bar R^{3}}\int_{\bar{\Gamma}(\bar t)}\bar \gamma_{\bar{s}}(\mathbf{\bar x}')\frac{(\bar{\textbf{x}}'-\bar{\textbf{x}})^{\perp}\cdot{\textbf{n}(\mathbf{\bar x})}}{|\bar{\textbf{x}}'-\bar{\textbf{x}}|^2}d\bar{s}(\mathbf{\bar x}').
\label{Vblp}
\end{equation}
Then the interface evolution in the scaled frame is given by Eq.~(\ref{evolvesc}).

\end{document}